\documentclass[aps,superscriptaddress,twocolumn,twoside,floatfix,pra,nofootinbib,a4paper]{revtex4-1}
\usepackage{times}
\usepackage{epsfig}
\usepackage{amsfonts}
\usepackage{amsmath}
\usepackage{amssymb}
\usepackage{amsthm}
\usepackage{color}
\usepackage{multirow}
\usepackage[normalem]{ulem}
\newcommand{\stkout}[1]{\ifmmode\text{\sout{\ensuremath{#1}}}\else\sout{#1}\fi}
\usepackage{latexsym}
\usepackage{mathrsfs}
\usepackage{natbib}
\usepackage{verbatim}
\usepackage[T1]{fontenc}
\usepackage{float}
\usepackage{subfigure}

\usepackage{graphicx}
\usepackage{xcolor}

\usepackage[colorlinks=true,linkcolor=blue,citecolor=magenta,urlcolor=blue]{hyperref}

\DeclareMathOperator{\Tr}{tr}

\newcommand{\ket}[1]{|#1\rangle}

\newcommand{\bracket}[3]{\langle#1|#2|#3\rangle}
\newcommand{\ketbra}[2]{|#1\rangle\langle#2|}

\newcommand{\norm}[1]{\left\lVert#1\right\rVert}

\begin{document}


\title{Nonlocality, steering and quantum state tomography in a single experiment}


\author{Chang-Jiang Huang}
\affiliation{Key Laboratory of Quantum Information, University of Science and Technology of China, CAS, Hefei 230026, China}
\affiliation{CAS Center for Excellence in Quantum Information and Quantum Physics, Hefei 230026, China}

\author{Guo-Yong~Xiang}
\email{gyxiang@ustc.edu.cn}
\affiliation{Key Laboratory of Quantum Information, University of Science and Technology of China, CAS, Hefei 230026, China}
\affiliation{CAS Center for Excellence in Quantum Information and Quantum Physics, Hefei 230026, China}

\author{Yu Guo}
\affiliation{Key Laboratory of Quantum Information, University of Science and Technology of China, CAS, Hefei 230026, China}
\affiliation{CAS Center for Excellence in Quantum Information and Quantum Physics, Hefei 230026, China}

\author{Kang-Da Wu}
\affiliation{Key Laboratory of Quantum Information, University of Science and Technology of China, CAS, Hefei 230026, China}
\affiliation{CAS Center for Excellence in Quantum Information and Quantum Physics, Hefei 230026, China}

\author{Bi-Heng Liu}
\email{bhliu@ustc.edu.cn}
\affiliation{Key Laboratory of Quantum Information, University of Science and Technology of China, CAS, Hefei 230026, China}
\affiliation{CAS Center for Excellence in Quantum Information and Quantum Physics, Hefei 230026, China}

\author{Chuan-Feng Li}
\affiliation{Key Laboratory of Quantum Information, University of Science and Technology of China, CAS, Hefei 230026, China}
\affiliation{CAS Center for Excellence in Quantum Information and Quantum Physics, Hefei 230026, China}

\author{Guang-Can Guo}
\affiliation{Key Laboratory of Quantum Information, University of Science and Technology of China, CAS, Hefei 230026, China}
\affiliation{CAS Center for Excellence in Quantum Information and Quantum Physics, Hefei 230026, China}

\author{Armin Tavakoli}
\email{armin.tavakoli@oeaw.ac.at}
\affiliation{Institute for Quantum Optics and Quantum Information - IQOQI Vienna, Austrian Academy of Sciences, Boltzmanngasse 3, 1090 Vienna, Austria}
\affiliation{D\'epartement de Physique Appliqu\'ee, Universit\'e de Gen\`eve, CH-1211 Gen\`eve, Switzerland}

\begin{abstract}			
We investigate whether paradigmatic measurements for quantum state tomography, namely mutually unbiased bases and symmetric informationally complete measurements, can be employed to certify quantum correlations. For this purpose, we identify a simple and noise-robust correlation witness for entanglement detection, steering and nonlocality that can be evaluated based on the outcome statistics obtained in the tomography experiment. This allows us to perform state tomography on entangled qutrits, a test of Einstein-Podolsky-Rosen steering and a Bell inequality test, all within a single experiment.  We also investigate the trade-off between quantum correlations and subsets of tomographically complete measurements as well as the quantification of entanglement in the different scenarios. Finally, we perform a photonics experiment in which we demonstrate quantum correlations under these flexible assumptions, namely with both parties trusted, one party untrusted  and both parties untrusted.

\end{abstract}
\date{\today}
\maketitle


%


\textit{Introduction.---} When Alice and Bob perform local measurements on a shared two-particle quantum state, they can observe correlations that go beyond classical models. These correlations are crucial for fundamental and applied aspects of quantum information science, but their precise sense of nonclassicality hinges on the assumptions made on the measurement devices. Three seminal scenarios have been widely studied in the literature: (i) both measurement devices are trusted, (ii) one device is trusted and the other is untrusted, (iii) both devices are untrusted. Nonclassicality in these scenarios correspond to detection of entanglement \cite{Guhne2009}, steering  \cite{Uola2020} and nonlocality \cite{Brunner2014} respectively. These are the paradigmatic resources for standard \cite{Horodecki2009}, one-sided device-independent \cite{Branciard2012} and fully device-independent \cite{Pironio2010} quantum information processing, respectively.

Therefore, it is important to experimentally classify states as entangled \cite{Friis2019}, steerable \cite{Bennet2012, Zeng2018, guo2019experimental, Designolle2020} or nonlocal \cite{ShalmLoopholefree, ZeilingerLoopholeFree2015, HansonLoopholeFree2015}. Typically, this requires three different experiments because the measurements tailored for detecting one form of nonclassicality generally are not well-suited for detecting another form of nonclassicality. For instance, the measurements optimal for violating the celebrated CHSH Bell inequality do not detect the steerability of any two-qubit isotropic state that is not already nonlocal \cite{Cavalcanti2015}. Another example is that correlations obtained from an optimal entanglement witness often admit a local-variable model when the measurement devices are no longer trusted.

Focusing on two-qutrit systems, we develop a correlation witness that detects quantum correlations in all three scenarios (i-iii) simultaneously. Our nonclassicality criteria are strongly noise tolerant and can be implemented using a flexible number of measurements on each side. We experimentally demonstrate their usefulness by detecting entanglement, steering and nonlocality for increasingly noisy states in a single experiment. Moreover, we show that the same data also can be used to deduce quantitative bounds on the entanglement in scenarios (i-iii).

Our protocol is based on  mutually unbiased bases (MUBs) \cite{Durt2010} and symmetric informationally complete measurements (SICs) \cite{Zauner, Renes2004}. Two bases are said to be MUBs if the magnitude of the overlap between any two basis elements chosen from different bases is constant. A set of $d+1$ MUBs in dimension $d$ is tomographically complete \cite{Wootters1989}. Similarly, a tomographically complete set of $d^2$ pure states is called a SIC if the overlap between any two states has the same magnitude. MUBs and SICs are conceptually related \cite{Wootters2006, Bengtsson2010, Bengtsson2012, Beneduci2013, TavakoliSICcompound}) and optimal for quantum state tomography \cite{Wootters1989, Scott2006, Rehacek2004}; which has prompted experimental demonstrations \cite{Adamson2010, Giovannini2013, Medendorp2011, Pimenta2013, Bent2015, hou2015, hou2018, tang2020}. 

In scenario (i), tomographically complete measurements enable full reconstruction of the quantum state and thus the evaluation of any entanglement criterion. Similarly,  in scenario (ii), such measurements are important for local reconstruction of the state assemblages, enabling optimal steering criteria. However,  beyond some notable special cases based on qubits \cite{Gisin2007, Plato}, there are no known Bell inequalities for which both complete sets of MUBs and SICs are optimal choices of measurements. Here, we present such a Bell inequality.

\textit{State tomography via MUBs and SIC.---} Let Alice and Bob share an unknown three-dimensional bipartite system $\rho$ whose density matrix they wish to estimate. To this end, they perform state tomography \cite{Leonhardt1995, TomoBook}. Alice performs one of nine dichotomic measurements  which respectively correspond to a projection onto each of the (unnormalised) states in the following SIC:
\begin{equation}\label{SIC}
\begin{Bmatrix}
& (0,1,-1) & (0,1,-\omega) & (0,1,-\omega^2)\\
& (-1,0,1) & (-\omega,0,1) & (-\omega^2,0,1)\\
& (1,-1,0) & (1,-\omega,0) & (1,-\omega^2,0)
\end{Bmatrix},
\end{equation}
where $\omega=e^{\frac{2\pi i}{3}}$.  A successful projection corresponds to the outcome $a=1$ and otherwise to the outcome $a=\perp$. We label the measurements by a pair of trits, $x_0,x_1\in\{0,1,2\}$, where $x_0$ labels the row and $x_1$ labels the column in \eqref{SIC}. Equivalently, we can use the label $x=3x_0+x_1+1$. Bob measures in one of the four  (unnormalised) MUBs
\begin{align}\label{MUBs}
&\begin{Bmatrix}
1 & 0 & 0\\
0 & 1 & 0\\
0 & 0 & 1
\end{Bmatrix}
& \begin{Bmatrix}
1 & 1 & 1 \\
1 & \omega & \omega^2 \\
1 & \omega^2 & \omega
\end{Bmatrix}
&& \begin{Bmatrix}
1 & \omega & \omega\\
\omega & 1 & \omega\\
\omega & \omega & 1
\end{Bmatrix}
&&\begin{Bmatrix}
1 & \omega^2 & \omega^2\\
\omega^2 & 1 & \omega^2\\
\omega^2 & \omega^2 & 1
\end{Bmatrix},
\end{align}
where the columns represent the basis states. The columns in a given basis thus represent his outcome, labelled $b=0,1,2$. The measurement basis is labelled by two bits, $y_0,y_1\in\{0,1\}$, which we equivalently write as $y=2y_0+y_1+1$. The probabilities $p(a,b|x,y)$ can be used to reconstruct the state $\rho$ by standard methods (see e.g.~Ref.~\cite{Schmied2016}).

\textit{Correlation witness.---} First noticed already in 1844 \cite{Hesse1844}, the SIC \eqref{SIC} and the MUBs \eqref{MUBs} exhibit a remarkable geometric relationship (see also Ref.~\cite{Bengtsson2012}). If a system is prepared in any of the nine states \eqref{SIC} and subsequently measured in any of the four bases \eqref{MUBs}, the probability distribution over the three possible outcomes admits a simple structure:  one outcome can never occur and the other two are equiprobable.  The events $(x,y,b)$ that cannot occur are characterised by the relation  $x_{y_1}=b \mod{3}$ when $y_0=0$ and $x_0+(-1)^{y_1}x_1=b \mod{3}$ when $y_0=1$. We define  $c_{xyb}=-1$ when this relation is satisfied and $c_{xyb}=+1$ otherwise. This observation can be extended to entangled system. If the shared state is maximally entangled, a local projection onto any of the elements in the SIC remotely prepares the other system in the same state (up to transpose). Inspired by this, we define a simple witness of the observed correlations:
\begin{align}\label{witness}
\mathcal{B}\equiv \sum_{x,y,b} c_{xyb}p(a=1,b|x,y)-2\sum_{x} p(a=1|x).
\end{align}

Consider now that Alice and Bob share the maximally entangled state $\ket{\phi_\text{max}}=\frac{1}{\sqrt{3}}(|00\rangle+|11\rangle+|22\rangle)$ exposed to a degree of isotropic noise: $\rho^{\text{iso}}_v=v\ketbra{\phi_\text{max}}{\phi_\text{max}} +\frac{1-v}{9}\openone$, where $v\in[0,1]$ is  the visibility. If Alice and Bob perform the measurements \eqref{SIC} and \eqref{MUBs} respectively, one obtains $\mathcal{B}=8v-2$, which in the noise-free case ($v=1$) becomes  $\mathcal{B}=6$. In what follows, we show that these quantum correlations detect nonclassicality in the scenarios (i-iii).

\textit{Correlation inequalities.---}  Based on the witness $\mathcal{B}$, we construct a Bell inequality, a steering inequality and an entanglement witness. For the first, we distrust Alice's and Bob's devices and determine the largest value of  $\mathcal{B}$ compatible with a local-hidden-variable (LHV) model: $p(a,b|x,y)=\sum_{\lambda}p(\lambda)p(a|x,\lambda)p(b|y,\lambda)$, where $\lambda$ is the local variable \cite{Brunner2014}. By considering all vertices of the polytope formed by these correlations \cite{Fine1982}, one finds a tight Bell inequality. For the second, we distrust only (say) Alice's device and assume that Bob performs the measurements \eqref{MUBs}. The (unnormalised) local states of Bob after Alice's measurements are given by the assemblage $\sigma_{a|x}=\Tr_\text{A}\left(A_{a|x}\otimes \openone \rho\right)$, where $A_{a|x}$ are Alice's measurement operators. We determine the largest value of $\mathcal{B}$ compatible with a local-hidden-state (LHS) model \cite{Wiseman2007}: $\sigma_{a|x}=\sum_{\lambda} p(\lambda)p(a|x,\lambda)\sigma_\lambda$ where $\sigma_\lambda$ are quantum states. This is achieved through a semidefinite program (see SM). For the third, we trust both devices and determine the largest value of $\mathcal{B}$ compatible with a separable (SEP) state: $\rho=\sum_\lambda p(\lambda) \psi_\lambda\otimes \phi_\lambda$, where $\psi_\lambda$ and $\phi_\lambda$ are pure qutrit states. This can be achieved by relaxing the set of separable states to the set of states with a positive partial transpose \cite{Horodecki1996} and then evaluating a semidefinite program (see SM). Then, we have saturated the resulting bound with a separable state to ensure tightness. 

In this manner, we obtain the tight correlation inequalities
\begin{equation}\label{ineqs}
\mathcal{B}\quad\stackrel{\text{SEP}}{\leq}\quad 0 \quad\stackrel{\text{LHS}}{\leq} \quad 2 \quad \stackrel{\text{LHV}}{\leq}\quad  4 \quad \stackrel{\text{Quantum}}{\leq} \quad 6. 
\end{equation}
We see that gradually relaxing the assumptions on Alice's and Bob's devices leads to stronger correlations in the relevant classical models. The last inequality (proven in SM) is a bound respected by quantum correlations for arbitrary states and local measurements. Note that this coincides with the value obtained from the SIC and MUBs.

The Bell inequality in \eqref{ineqs} is violated whenever $v>\frac{3}{4}$. No better Bell inequality is possible for $p(a,b|x,y)$  because for any smaller $v$ the full distribution can be simulated in a local-variable model. 
To further explore the role of tomographically complete measurements in our Bell inequality test, we have numerically searched for the largest value of $\mathcal{B}$ possible with real Hilbert spaces of a fixed dimension. For qutrits, we found at best $\mathcal{B}\approx 5.469$\footnote{By increasing the Hilbert space dimension to five, we could still not find a maximal violation in real Hilbert spaces ($\mathcal{B}\approx 5.731$). In dimension six, one can achieve $\mathcal{B}=6$ in real spaces.}. This illustrates the need for exploiting the full structure of qutrit Hilbert space.

If the isotropic state is too noisy ($v<\frac{3}{4}$) to violate the Bell inequality, we can instead violate the steering inequality in \eqref{ineqs} if $v>\frac{1}{2}$. Interestingly, the critical visibility can be further reduced by fine-graining the witness. In our quantum protocol, the marginal distribution of Alice is $p(a=1|x)=\frac{1}{3}$ regardless of the visibility $v$. Hence, the second term in \eqref{witness} is $t\equiv 2\sum_{x} p(a=1|x)=6$. If we fix $t=6$,  the analysis of the quantum protocol remains unchanged but the bound for LHS  models is reduced to $\mathcal{B}\lesssim 1.759$. This allows us to detect steering for $v\gtrsim 0.470$. Importantly, this threshold cannot be further reduced by using any other witness since it equals that obtained from considering full tomography of $\sigma_{a|x}$. Notably, the visibility threshold  is lower than that encountered for a two-qubit isotropic state under general projective measurements ($v>\frac{1}{2}$) \cite{Wiseman2007}. Furthermore, it is not far from the threshold limiting the most general experiment based on any number of dichotomic projective measurements, namely $v\gtrsim 0.423$ \cite{Nguyen2020, Wiseman2007}. Finally, we have also considered the reverse situation in which Alice is trusted and Bob is untrusted. Then, the witness detects steering for $v\gtrsim 0.482$ (see also Ref.~\cite{Bavaresco2017}).

If the isotropic state is too noisy to violate the steering inequality, we can instead violate the separable bound in \eqref{ineqs} if $v>\frac{1}{4}$. Note that this is optimal, because below this visibility, the isotropic state is separable  \cite{Horodecki1996}. 

The above analysis, based on $\mathcal{B}$, provides a simple criterion that applies equally well to detecting three forms of nonclassicality in manner that is optimal for the isotropic state in our input/output scenario. However, it is often relevant to detect quantum correlations in a more resource-efficient way; by using fewer measurements. Next, we go beyond the inequalities \eqref{ineqs} and consider quantum correlations in the full statistics $p(a,b|x,y)$ when the number of settings on Alice and Bob is a degree of freedom.

\textit{Correlations from fewer settings.---} We investigate the trade-off between efficient nonclassicality criteria (using few measurements) and stronger criteria (using more measurements). For this purpose, a natural setting is to let Alice and Bob measure only subsets of the elements in the SIC and MUBs respectively, i.e.~to use only settings $x\in\{1,\ldots,n\}$ and $y\in\{1,\ldots,m\}$ for some $2\leq n\leq 9$ and $2\leq m\leq 4$. If small choices of $(n,m)$ are insufficient to detect the nonclassicality of a state, additional elements of the SIC and MUBs can be measured, thus step-by-step approaching a tomographically complete set \cite{Bae2019}. In SM, we employ convex programming methods to determine visibility thresholds for the state $\rho_v^\text{iso}$ above which the probability distribution $p(a,b|x,y)$, for given $(n,m)$, implies entanglement, steering and nonlocality respectively. In particular, we find that entanglement is detected already in the minimal scenario, $(n,m)=(2,2)$, albeit at high visibility. We illustrate in Table~\ref{Tablefewer} the results when $m=4$ and Alice performs only $n$ measurements. For entanglement, the detection is optimal already at $n=6$. Steering is robustly detected already at $n=2$. However, nonlocality requires at least $n=5$.

\begin{table}[t!]
	\begin{tabular}{c|cccccccc}
		$n$    & 2    & 3    & 4    & 5    & 6    & 7    & 8    & 9    \\ \hline
	SEP & 0.63 & 0.40 &0.36 & 0.31 & 0.25 & 0.25 & 0.25 & 0.25 \\
		LHS                         & 0.73                         & 0.67                         & 0.60                         & 0.56 & 0.53 & 0.51 & 0.49 & 0.47 \\
		LHV                         & -                            & -                            & -                            & 0.91 & 0.86 & 0.86 & 0.81 & 0.75
	\end{tabular}
\caption{Visibility thresholds for entanglement, steering and nonlocality based on $p(a,b|x,y)$ when Bob measures the four MUBs and Alice measures $n$ elements of the SIC on the state $\rho^\text{iso}_v$. The symbol ``-'' indicates that an LHV model is possible.}\label{Tablefewer}
\end{table}

\textit{Entanglement quantification.---} Our above results classify a state as either entangled, steerable or nonlocal. We now go further and show that entanglement also can be quantified, specifically through the entanglement negativity \cite{Vidal2002},  based on the observed quantum correlations, in a  standard, one-sided device-independent and fully device-independent manner respectively.  The negativity is defined as $N(\rho)=\frac{1}{2}\left(\norm{\rho^{T_\text{A}}}_1-1\right)$, where $T_\text{A}$ denotes partial transpose. It can alternatively be computed as the semidefinite program $N(\rho)=\min\{\Tr\left(\sigma_-\right)| \rho=\sigma_+-\sigma_-, \sigma_\pm^{T_\text{A}}\geq 0\}$.

In the trusted scenario (i), we consider the case of $m=4$ and $n=6$, which we previously found was the minimal setting for optimal detection of the isotropic state. For given $v$, we solve the semidefinite program for $N$ under the additional constraint that all probabilities $p(a,b|x,y)$ are reproduced by the state $\rho$ when applying the SIC and MUBs.  For  $\frac{1}{4}\leq v\lesssim 0.542$, we find the bound $N\geq \frac{8v-2}{9}$. When $ v\gtrsim 0.542$, our numerical bound can be approximated up to  the third decimal by the polynomial $N\gtrsim 0.130 v^2 + 1.418 v - 0.548$. A non-trivial amount of entanglement is quantified for every nontrivial visibility. 

\begin{figure}
	\centering
	\includegraphics[width=\columnwidth]{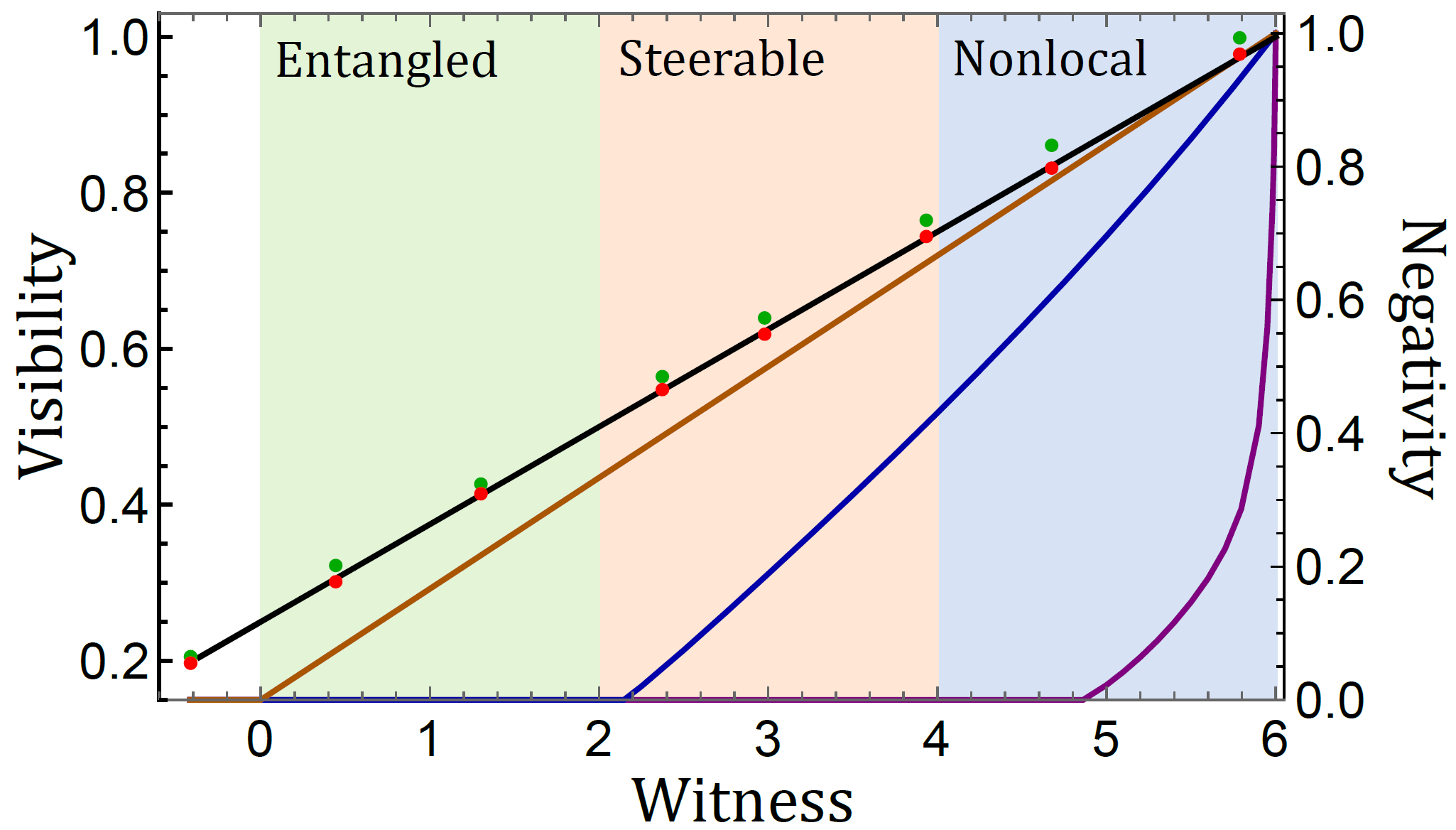}
	\caption{(Left-bottom axes) Experimentally measured value of witness $\mathcal{B}$ for isotropic states with eight different target visibilities (green points). The red points represent the same value of $\mathcal{B}$ but with the visibility estimated from tomography. The theoretical prediction is represented by the black line. In the blue, orange and green regions, our data detects nonlocality, steering and entanglement respectively via the inequalities \eqref{ineqs}. (Right-bottom axes) The purple, blue and orange curves represent  lower bounds on the entanglement negativity as a function of the witness $\mathcal{B}$ in the device-independent, one-sided device-independent and trusted scenario respectively.}\label{FigResults}
\end{figure}

In the one-sided device-independent scenario (ii) and in the fully device-independent scenario (iii), we quantify entanglement based on the parameter $\mathcal{B}$ used in the steering and Bell inequalities in \eqref{ineqs}. This can be achieved using the semidefinite relaxation hierarchies in Ref.~\cite{Pusey2013} and Ref.~\cite{Moroder2013} respectively. By using a level intermediate between the first and second, we have obtained non-trivial bounds on $N$. These are displayed in Figure~\ref{FigResults}. For completeness, we also illustrate in Figure~\ref{FigResults} the bound  $N\geq \frac{\mathcal{B}}{6}$ for the trusted scenario (i) using tomographically complete measurements.  For the ideal case of $\mathcal{B}=6$, all our numerical bounds return the optimal negativity $N=1$. However, below $\mathcal{B}\approx 2.14$ (steering) and $\mathcal{B}\approx 4.86$ (nonlocality) respectively, we have $N=0$, which is suboptimal. The reason is that since we work with enough settings for tomographic completeness, and go beyond a qubit scenario, it is computationally expensive to consider semidefinite relaxations sufficiently large to obtain nearly-optimal bounds. It is likely that already the evaluation of the second relaxation level will improve the bounds for scenarios (ii) and (iii).

\begin{figure}[t!]
	\begin{center}
		\includegraphics [width=0.95\columnwidth]{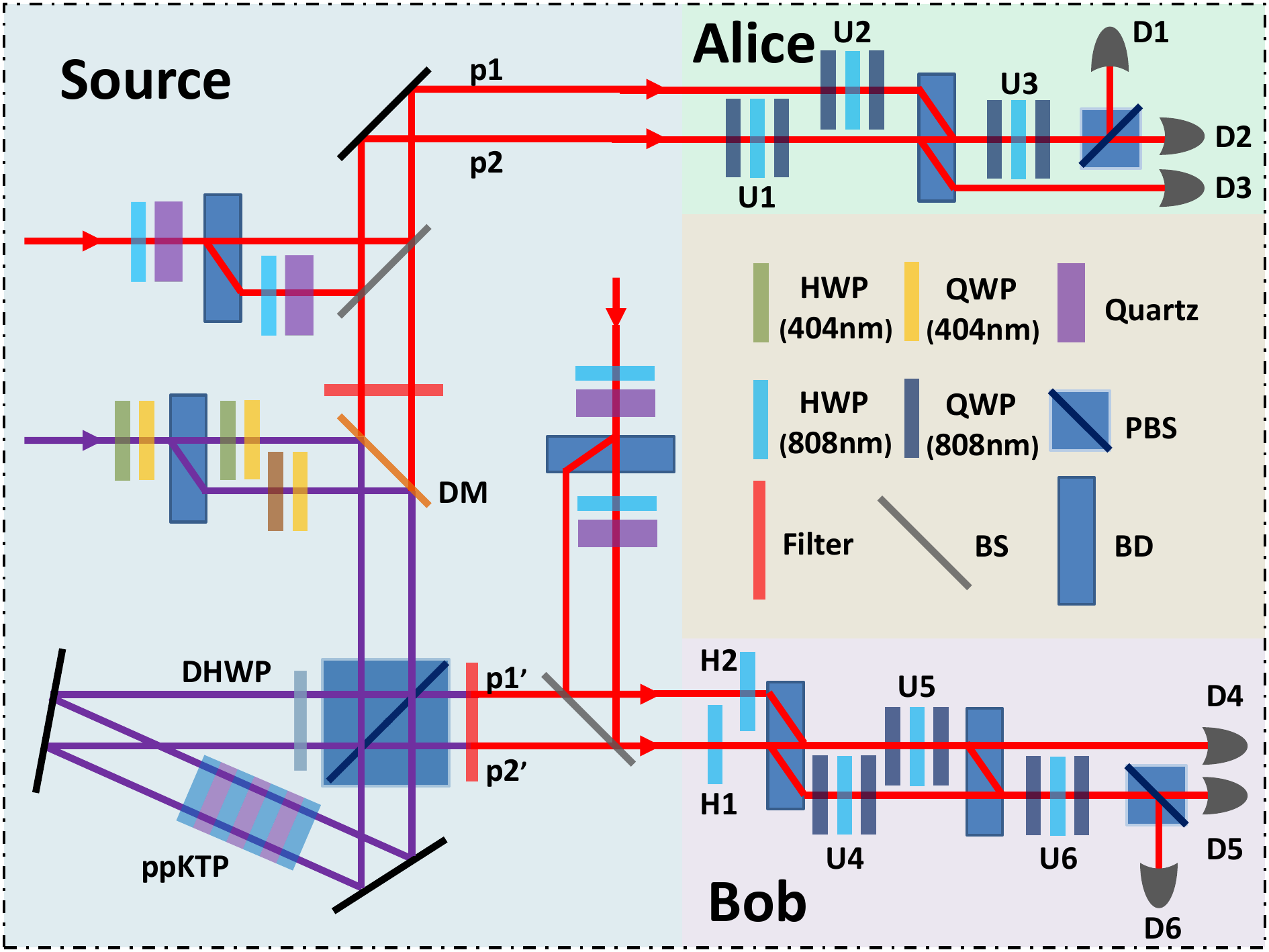}
	\end{center}
	\caption{Experimental setup. 
	In the state preparation, a continuous wave violet laser at 404 nm is used to generate a pair of entangled photons via a type-II phase-matched SPDC process in a Sagnac structure. The isotropic state  is encoded by both path and polarisation degrees of freedom of photons and its visibility is controlled via the configuration composed of two HWPs, two quartz crystals and a BD. Alice's measurements are performed using a series of HWPs, BDs, QWPs and a PBS.  Bob's measurements are performed using two HWPs, three universal unitary qubit gates, two BDs and a PBS.}
	\label{Fig:setup}
\end{figure}

\textit{Experimental setup.---} We have demonstrated nonlocality, steering and tomography in a photonics experiment. The setup (see Fig.~\ref{Fig:setup}) is composed of three parts. The first part is the preparation of the isotropic state $\rho_v^\text{iso}$, which is encoded in both the polarisation and the path degree of freedom of the photons.  The horisontal-polarisation (vertical-polarisation) photon taking path $p_1$ ($p'_1$) encodes the state $|0\rangle$, the vertical-polarisation (horisontal-polarisation) photon taking path $p_2$ ($p'_2$) encodes the state $|1\rangle$ and the horisontal-polarisation (vertical-polarisation) photon passing path $p_2$ ($p_2'$) encodes the state $|2\rangle$. The maximally entangled state $|\phi_\text{max}\rangle$ is generated through the type-II phase-matched spontaneous parametric down-conversion (SPDC) process in the Sagnac structure \cite{guo2019experimental}. The visibility parameter is controlled by inserting quartz crystals to destroy the coherence of a photon-pair state $\frac{1}{3}(|0\rangle+|1\rangle+|2\rangle)\bigotimes(|0\rangle+|1\rangle+|2\rangle)$ \cite{verbanis2016resource}.

The second part is the realisation of Alice's measurements (see Fig.~\ref{Fig:setup}). Alice can choose between performing one of nine different binary-outcome measurements, corresponding to the SIC \eqref{SIC}, on her qutrit. These measurements are realised by three half-wave plates (HWPs), six quarter-wave plates (QWPs), one beam displacer (BD), a polarisation beam splitter (PBS) and three single photon detectors. The detector $D1$ is used to record the outcome  $a=1$ and the detectors $D2$ and $D3$ record the outcome $a=\perp$. 

The third part is the realisation of Bob's measurements (see Fig.~\ref{Fig:setup}). The four MUBs \eqref{MUBs} are realised using five HWPs, six QWP, two BDs, a PBS and three single photon detectors. The detector $D4$, $D5$ and $D6$ are used to record outcomes $b=0,1,2$ respectively. The wave-plate angles corresponding to Alice's and Bob's measurements are given in SM.

\textit{Experimental results.---} We have realised the isotropic state for eight different target visibilities. For each case, we perform the measurements \eqref{SIC} and \eqref{MUBs} with approximately $10^5$ shots per setting $(x,y)$. In SM we use the relative frequencies to estimate the density matrix via the maximum-likelihood method \cite{jevzek2003quantum}. The visibility of the reconstructed matrix is, however, somewhat smaller than our targeted visibility (see SM), primarily due to imperfections in the source and  the interference between the BDs. For instance, when we target the preparation of the maximally entangled state ($v=1$), the visibility of the reconstructed state is approximately $0.977$.

In Fig.~\ref{FigResults}, we present the measured value of the witness $\mathcal{B}$ for the different visibilities; green (red) points are target (estimated) visibilities.  The standard deviations are all of a magnitude smaller than $2\times 10^{-2}$, corresponding to accumulated statistical and systematic errors. Systematic errors include that our coincidence rate is approximately $1000$ per second of which roughly one is a dark count, that the wave-plates are subject to a small offset ($<0.5\deg$) and induce a phase error ($<1.2\deg$). The solid black line represents the theoretical witness value $\mathcal{B}=8v-2$. We see that our data accurately confirms the theoretical prediction. When the visibility is high we observe that the data violates the Bell inequality in Eq.~\eqref{ineqs}. To this end, we assume fair-sampling in the detection events. When the visibility is further reduced, the data violates the steering inequality in Eq.~\eqref{ineqs}. For even smaller visibilities, we still detect entanglement through the witness in Eq.~\eqref{ineqs}. For our final data point (with target visibility $v\approx 0.205$), the isotropic state is separable.  For instance, when no noise is deliberately added to the system (target $v=1$), we observe $\mathcal{B}=5.796\pm  0.015$. This certifies a negativity of $N=0.940$ in the steering scenario, which is only marginally worse than $N=0.969$ computed from the reconstructed density matrix. In the Bell scenario, our value of $\mathcal{B}$ certifies at least $N=0.287$. Also, the witness value exceeds the largest value found using quantum theory constrained to real three-dimensional systems and real five-dimensional systems by $21.8$ and $4.3$ standard deviations respectively.

\textit{Discussion.---} We have performed quantum state tomography on increasingly noisy entangled states via mutually unbiased bases and  symmetric informationally complete projections. Using a simple correlation witness, we showed that the outcome statistics can be re-interpreted both as a demonstration of steering and a demonstration of nonlocality. Moreover, we showed how classification of nonclassical properties can be achieved using a tunable number of measurements and that entanglement can be quantified in the three scenarios using a single setup.

Our work, together with the foundational interest in Bell inequalities tailored for important quantum measurements \cite{Kaniewski2019, Tavakoli2019}, motivates a natural open problem: to generalise our findings to MUBs and SICs of any dimension. It interesting to investigate how the noise-tolerance of such a generalisation scales with the system dimension and the choice of assumption. For practical purposes, this is particularly relevant for steering since witnesses of our type only use binary-outcome measurements for the untrusted party (Alice). This facilitates experimental implementation while simultaneously, e.g.~for the isotropic state, does not imply a significant fundamental constraint on the detectable states \cite{Nguyen2020}.

Finally, another interesting continuation of our work is to develop noise-robust correlation witnesses for detection of nonlocality, steering and entanglement in quantum networks \cite{Review}. In such networks, in analogy with our approach here, it is known that measurements in MUBs and entangled measurements based on SICs are natural candidates to reveal quantum nonlocality \cite{Gisin2019, Tavakoli2020, Baumer2020}.

\begin{acknowledgments}
We thank Roope Uola and J\k{e}drzej Kaniewski for comments and discussions. We are obliged to Ingemar Bengtsson for inspiration. This work was supported by the Swiss National Science Foundation (Starting grant DIAQ, NCCR-SwissMAP and Early PostDoc Mobility fellowship P2GEP2 194800), the National Natural Science Foundation of China under Grants (Nos. 11574291, 11774334, 11874345 and 11904357), the National Key Research and Development Program of China (No.2017YFA0304100, No.2018YFA0306400), Key Research Program of Frontier Sciences, CAS (No.QYZDY-SSW-SLH003), Science and Technological Fund of Anhui Province for Outstanding Youth (2008085J02).
\end{acknowledgments}

\bibliography{references_mubsic}

\appendix
\onecolumngrid

\section{Maximal quantum violation of the Bell inequality}\label{AppendixQmax}
We prove that $\mathcal{B}\leq 6$ in quantum theory. We denote Alice's outcome-one measurement operators by $A_x$ and Bob's measurement operators by $B_{b|y}$. Without loss of generality, we can restrict $A_x$ and $B_{b|y}$ to projective operators. First, we focus on just the first of the two  terms in the witness, which we denote $\mathcal{B}'$, which in a quantum model reads
\begin{equation}
\mathcal{B}'=\sum_{x,y,b} c_{xyb}\bracket{\psi}{A_x\otimes B_{b|y}}{\psi}.
\end{equation}
Using the Cauchy-Schwarz inequality, we obtain
\begin{align}\nonumber
\mathcal{B}'=\sum_x \bracket{\psi}{A_x\otimes \sum_{y,b}c_{xyb}B_{b|y}}{\psi} \leq \sum_x \sqrt{\bracket{\psi}{A_x^2\otimes \openone}{\psi}}\sqrt{\bracket{\psi}{\openone\otimes \left(\sum_{y,b}c_{xyb}B_{b|y}\right)^2}{\psi}}.
\end{align}
The expression under the first square-root is just the marginal probability $p(a=1|x)$. Let us denote it by $q_x\in[0,1]$. Then the Cauchy-Schwarz inequality  gives
\begin{equation}
\mathcal{B}'\leq \sqrt{\sum_x q_x}\sqrt{\sum_x\bracket{\psi}{\openone\otimes \left(\sum_{y,b}c_{xyb}B_{b|y}\right)^2}{\psi}}.
\end{equation}
Expanding the expression under the second square-root, one finds that it is independent of Bob's choice of measurements,
\begin{equation}
\sum_x \left(\sum_{y,b}c_{xyb}B_{b|y}\right)^2=48\openone.
\end{equation}
Using this fact, and the simplifying notation $q=\sum_x q_x$, we arrive at
\begin{equation}\label{step}
\mathcal{B}'\leq \sqrt{48q}.
\end{equation}
Turning to the full witness expression $\mathcal{B}$, we use Eq.~\eqref{step} to obtain
\begin{equation}\label{finstep}
\mathcal{B}\leq \sqrt{48q}-2q.
\end{equation}
In order to evaluate the maximal value of the right-hand-side, we solve $0=\frac{d}{dq}\left(\sqrt{48q}-2q\right)$ which gives $q=3$. Inserted into Eq.~\eqref{finstep}, we find
\begin{equation}
\mathcal{B}\leq 6.
\end{equation}

\newpage
\section{Nonlocality, steering and entanglement from subsets of SIC and MUBs}\label{AppPartial}
Consider that we perform only a subset of the measurements for quantum state tomography. Specifically, Alice performs only the first $n\in\{2,\ldots,9\}$ binary-outcome measurements corresponding to the SIC and Bob performs only the first $m\in\{2,3,4\}$ MUBs. The full tomography case therefore corresponds to $(n,m)=(9,4)$. How well can we detect nonlocality, steering and entanglement using only such subsets of the SIC and subsets of  the complete set of MUBs?

\subsection{Nonlocality}
We investigate nonlocality in the measured probabilities $p(a,b|x,y)$ when Alice only performs her first $n$ measurements and Bob performs the complete set of MUBs when sharing the isotropic state $\rho_v^\text{iso}$. We note that this analysis readily extends also to other states. We have used a linear program to evaluate the critical visibility for a simulation of $p(a,b|x,y)$ in a local model. This linear program is
\begin{align}\nonumber
&\qquad \qquad \qquad \qquad   \max_{q_{\lambda_1,\lambda_2}} v \\\nonumber
&\text{such that } \quad p(a,b|x,y)=\sum_{\lambda_1=1}^{2^n}\sum_{\lambda_2=1}^{3^m} q_{\lambda_1,\lambda_2} D_{\lambda_1}(a|x)D_{\lambda_2}(b|y),\\
& q_{\lambda_1,\lambda_2}\geq 0, \qquad \text{and} \qquad \sum_{\lambda_1=1}^{2^n}\sum_{\lambda_2=1}^{3^m} q_{\lambda_1,\lambda_2}=1,
\end{align}
where $D_{\lambda_1}(a|x)$ and $D_{\lambda_2}(b|y)$ are all deterministic functions from Alice's and Bob's respective inputs to their respective outputs. Note that the probability distribution itself is a function of $v$ since it is given by measurements on the isotropic state of visibility $v$. We have  evaluated this program for every pair $(n,m)$ and found that only when $m=4$ and $n\geq 5$ it is possible to detect nonlocality. The critical visibilities required for nonlocality are given in Table~\ref{TableBell}. 
\begin{table}[h!]
	\begin{tabular}{c|ccccc}
		n & 5      & 6      & 7      & 8      & 9    \\ \hline
		v & 0.914 & 0.857 & 0.857 & 0.813 & 0.750
	\end{tabular}
	\caption{Critical visibility of the isotropic state for a local model of $p(a,b|x,y)$ when Alice  performs the first $n$ binary-outcome measurements associated to the Hesse SIC and Bob performs the given complete set of MUBs.}\label{TableBell}
\end{table}

\subsection{Steering}
Consider that Alice is untrusted and only performs $n$ of the measurements in the SIC while Bob performs the four MUBs ($m=4$). Since Bob's measurements are tomographically complete, he can reconstruct the steering assemblage $\sigma_{a|x}=\Tr_\text{A}\left(A_{a|x}\otimes \openone \rho_v^\text{iso}\right)$. We can determine the critical visibility for steerability by solving the following semidefinite program:
	\begin{align}\nonumber\label{SDPsteering}
	&\qquad \qquad \qquad \qquad   \max_{\sigma_{\lambda}} v \\\nonumber
	&\text{such that } \quad \sigma_{a|x}=\sum_{\lambda=1}^{2^n} D_{\lambda}(a|x)\sigma_{\lambda},\\
	& \sigma_\lambda\geq 0, \qquad \text{and} \qquad \sum_{\lambda=1}^{2^n} \sigma_\lambda=\frac{\openone}{3},
	\end{align}
	where $D_\lambda(a|x)$ are all deterministic functions mapping Alice's input to her output and $\sigma_\lambda$ is a subnormalised state. That the final constraint in the above SDP stems from the fact that Bob's share of the isotropic state is $\frac{\openone}{3}$ regardless of $v$. Note that $v$ appears indirectly in the assemblages $\sigma_{a|x}$. Furthermore, we note that if this program certifies the state as steerable, one can always extract a steering inequality by considering the dual of the above SDP.

In Table~\ref{TableAliceToBob} we give the critical visiblity for detecting steering for the different values of $n$. Note that the case of $n=9$ coincides with the critical visibility given in the main text based on the quantity $\mathcal{B}$. We find that steering is detected for every $n$. In fact, even in the case of only using two of the nine SIC projections, the visibility is lower than that required to violate our Bell inequality.

\begin{table}[h!]
	\begin{tabular}{c|cccccccc}
		n & 2      & 3      & 4      & 5      & 6      & 7      & 8      & 9      \\ \hline
		v & 0.727 & 0.667 & 0.602 & 0.561 & 0.532 & 0.510 & 0.488 & 0.470
	\end{tabular}
	\caption{Critical visibility of isotropic state for steering when the untrusted party performs the first $n$ binary-outcome measurements associated to the Hesse SIC.}\label{TableAliceToBob}
\end{table}

The SDP in Eq.~\eqref{SDPsteering} can be straightforwardly adapted to the opposite case, in which Alice is trusted and Bob is not. Then, we consider that Alice performs local tomography using the full set of SIC measurements while Bob measures only $m$ MUBs. In Table~\ref{TableBobToAlice}, we give the critical visibilities for steerability. We see that even two MUBs are sufficient to demonstrate steering of the isotropic state at visibilities for which no Bell inequality violation is known.

\begin{table}[h!]
	\begin{tabular}{c|ccc}
		m & 2      & 3      & 4      \\ \hline
		v & 0.683 & 0.569 & 0.481
	\end{tabular}
	\caption{Critical visibility of isotropic state for steering when the untrusted party performs the first $m$ ternary-outcome measurements in the given complete set of MUBs.}\label{TableBobToAlice}
\end{table}

\subsection{Entanglement detection}
 We now consider entanglement detection from the probability distribution $p(a,b|x,y)$ based on the isotropic state and $n$ and $m$ measurement respectively for Alice and Bob. In contrast to the cases of nonlocality and steering, it is difficult to give a necessary and sufficient condition for whether there exists a separable model for the full distribution. We address the problem by relaxing the set of separable states to the set of quantum states with a positive partial transpose. This allows us to identify an upper bound for the critical visibility needed for entanglement detection through $p(a,b|x,y)$ by solving the following semidefinite program:
	\begin{align}\nonumber\label{SDPentanglement}
	&\qquad \qquad \qquad \qquad   \max_{\rho} v \\\nonumber
	&\text{such that } \quad p(a,b|x,y)=\Tr\left(A_{a|x}\otimes B_{b|y} \rho \right),\\
	& \rho\geq 0, \qquad \Tr\left(\rho\right)=1 \qquad  \text{and} \qquad \rho^{T_\text{A}}\geq 0.
	\end{align}	
	We have evaluated this SDP for every pair $(n,m)$ and the resulting upper bounds on the critical visibility for entanglement detection is given in Table~\ref{tabSep}. We see that for every pair $(n,m)$ it is possible to detect entanglement. Moreover, we see that the required visibility reduces as we increase the number of settings on Alice or Bob. In particular, for the isotropic state, optimal entanglement detection is achieved already for $m=4$ and $n=6$, which is more efficient than the criterion based on the quantity $\mathcal{B}$ given in the main text. 
	
	\begin{table}[h!]
		\begin{tabular}{c|cccccccc}
			m\textbackslash{}n & 2                              & 3                           & 4                           & 5                           & 6      & 7      & 8      & 9      \\ \hline
			2                  & 0.9132 & 0.8117 & 0.6553 & 0.5680 & 0.5000 & 0.5000 & 0.5000 & 0.5000 \\
			3                  & 0.7933 & 0.6667 & 0.5259 & 0.4288 & 0.3487 & 0.3448 & 0.3333 & 0.3333 \\
			4                  & 0.6323                         & 0.4000                      & 0.3633                      & 0.3136                      & 0.2500 & 0.2500 & 0.2500 & 0.2500
		\end{tabular}
		\caption{Upper bound on critical visibility of isotropic state for entanglement detection when Alice uses $n$ settings in the SIC and Bob uses $m$ MUBs.}\label{tabSep}
	\end{table}
	
	We note that by considering the  dual of the above SDP one can convert these certificates into entanglement witnesses (inequalities) that are more friendly to experimental implementation. Lastly, we remark that one could in principle hope to obtain tighter bounds on the critical visibiltiy than that provided by the relaxation \eqref{SDPentanglement}. This can be achieved using the DPS hierarchy, which is based on considering symmetric extensions of states with a positive partial transpose. The SDP \eqref{SDPentanglement} corresponds to the lowest level in this hierarchy. We have also implemented the second level of the hierarchy, based on a state $\rho_{AB_1B_2}$ which is invariant under permutation of Bob's two subsystems and which has a positive partial transpose with respect to Bob's systems.  However, for most choices of $(n,m)$ this does not improve the critical visibility and for the few cases in which and improvement is obtained, it is only a minor one.

\newpage
\section{Wave-plate angles}\label{AppAngles}

In Table~\ref{SICangles} we present the wave-plate angles for the nine measurements corresponding to the considered SIC. Similarly, the wave-plate angles corresponding to the four MUBs are given in  Table~\ref{MUBangles}. In particular, the angles for the unitary gate correspond to the QWP, HWP, QWP from the left to the right.

\begin{table}[htbp]
	\centering
	\setlength{\tabcolsep}{1.75mm}{
		\begin{tabular}{|c|ccc|ccc|ccc|}
			\hline
			SIC   & \multicolumn{3}{c|}{U1} & \multicolumn{3}{c|}{U2} & \multicolumn{3}{c|}{U3} \\
			\hline
			1     & -22.5 & 22.5  & -22.5 & 0     & 0     & 0     & 0     & 0     & 0 \\
			2     & -45   & 52.5  & 0     & 0     & 0     & 0     & 0     & 0     & 0 \\
			3     & 45    & 52.5  & 90    & 0     & 0     & 0     & 0     & 0     & 0 \\
			4     & 0     & 45    & 90    & 0     & 0     & 0     & -22.5 & 22.5  & -22.5 \\
			5     & 0     & 45    & 90    & 0     & 0     & 0     & -45   & 52.5  & 0 \\
			6     & 0     & 45    & 90    & 0     & 0     & 0     & 45    & 52.5  & 90 \\
			7     & 0     & 0     & 0     & 0     & 0     & 0     & -22.5 & 22.5  & -22.5 \\
			8     & 0     & 0     & 0     & 0     & 0     & 0     & 45    & 52.5  & 90 \\
			9     & 0     & 0     & 0     & 0     & 0     & 0     & -45   & 52.5  & 0 \\
			\hline
	\end{tabular}}
	\caption{The angles for Alice's nine measurements corresponding to a SIC.}
	\label{SICangles}%
\end{table}%

\begin{table}[htbp]
	\centering
	\setlength{\tabcolsep}{1.1mm}{
		\begin{tabular}{|c|c|c|ccc|ccc|ccc|}
			\hline
			MUB   & H1    & H2    & \multicolumn{3}{c|}{U4} & \multicolumn{3}{c|}{U5} & \multicolumn{3}{c|}{U6} \\
			\hline
			1     & 0     & -45   & 0     & -45   & 90    & 45    & 90    & -45   & 0     & -45   & 90 \\
			2     & 67.5  & 45    & 0     & 45    & 0     & 27.37 & -17.63 & 27.37 & 45    & 0     & 90 \\
			3     & 67.5  & 45    & 0     & 45    & 0     & 45    & 12.37 & 9.74  & -45   & -60   & 0 \\
			4     & 67.5  & 45    & 0     & 45    & 0     & 45    & -57.37 & 80.26 & 45    & 52.5  & 90 \\
			\hline
	\end{tabular}}
	\caption{The angles for Bob's four MUB measurements}
	\label{MUBangles}
\end{table}%

\newpage
\section{Experimental  tomography of prepared states}\label{AppendixTomography}
We have prepared the isotropic state for eight different target visibilities and reconstructed the density matrix using a maximum-likelihood estimate. Here, we display the results of the tomographic procedure for all eight cases (red for real part of the density matrix, blue for imaginary part of the density matrix).
\begin{figure}[htbp]
	\begin{center}
		\subfigure(a){\includegraphics [width=8.77cm,height=5.7cm]{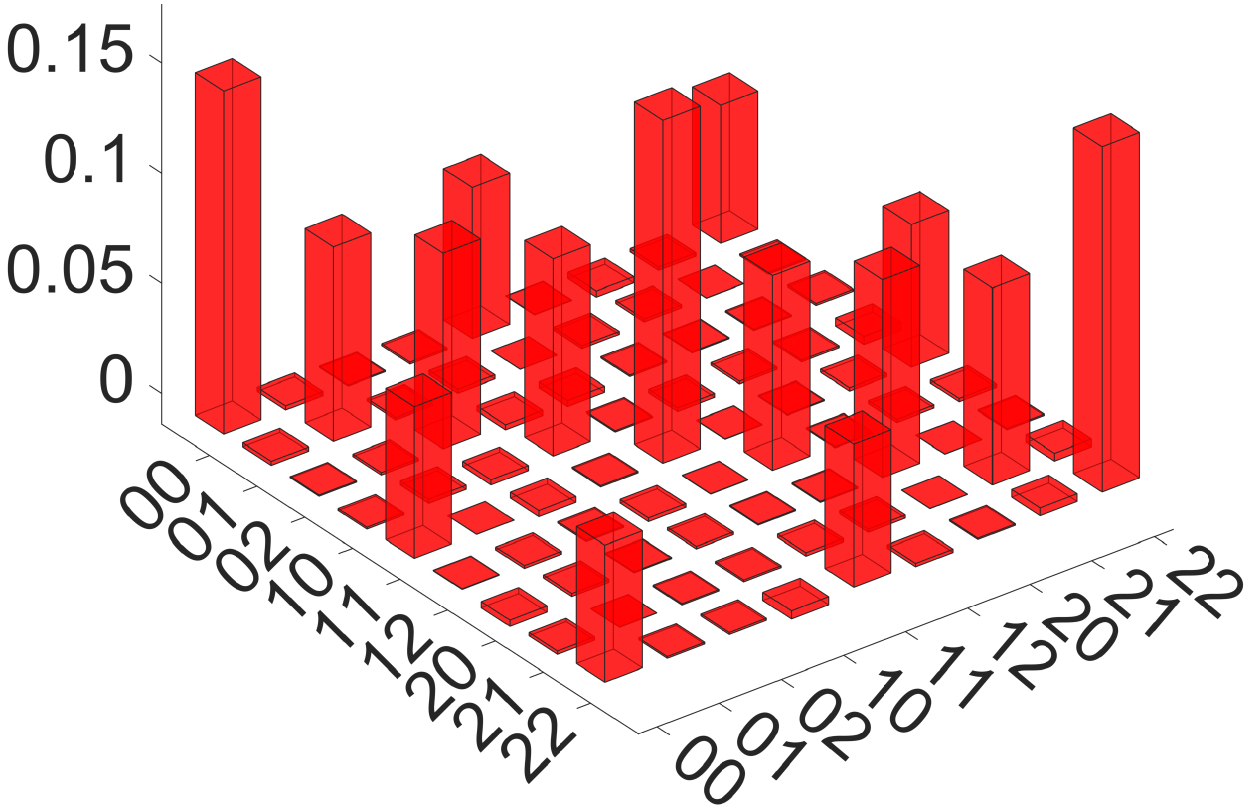}}
		\subfigure(b){\includegraphics [width=8.35cm,height=5.7cm]{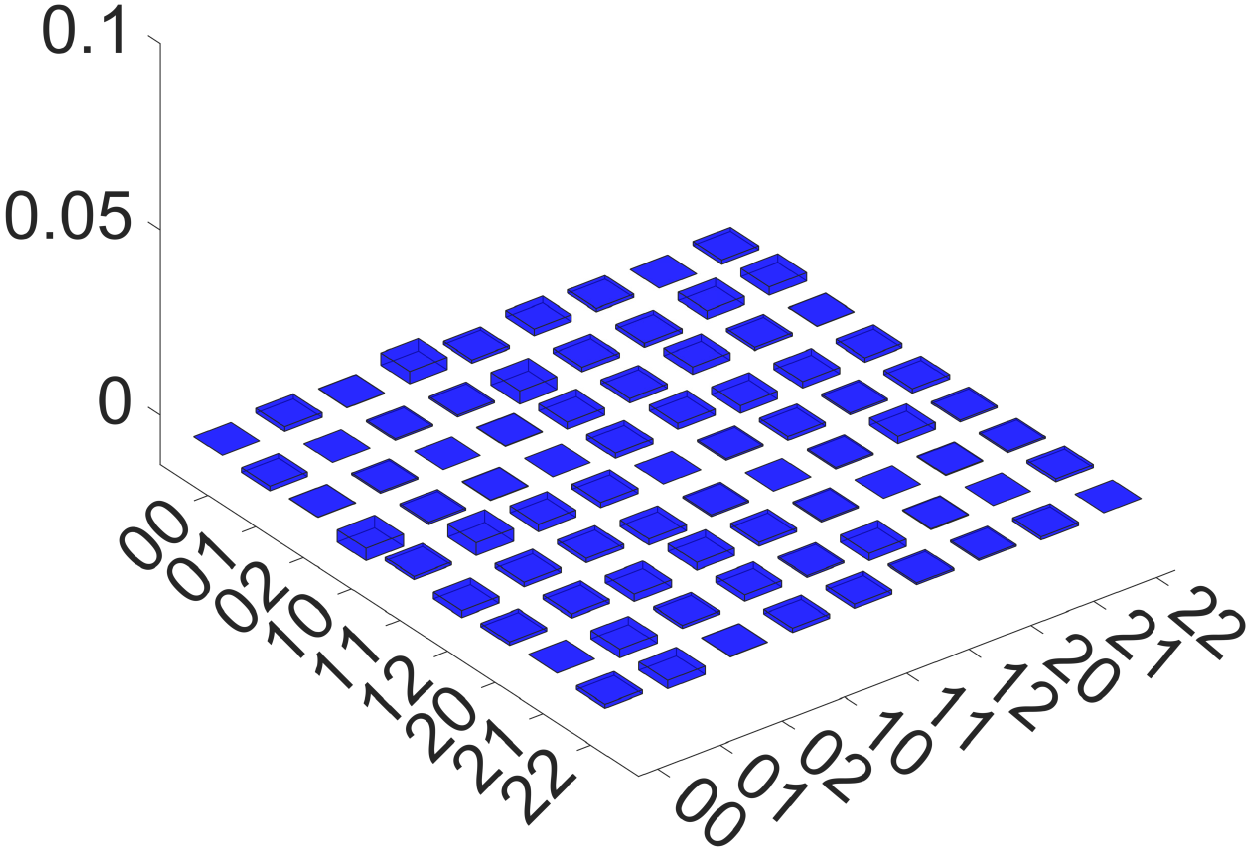}}
	\end{center}
	\caption{Target $v=0.205$, estimate $v=0.197$. (a) The real elements. (b) The imaginary elements.}
	\label{Fig:tomography1}
\end{figure}

\begin{figure}[htbp]
	\begin{center}
		\subfigure(a){\includegraphics [width=8.77cm,height=5.7cm]{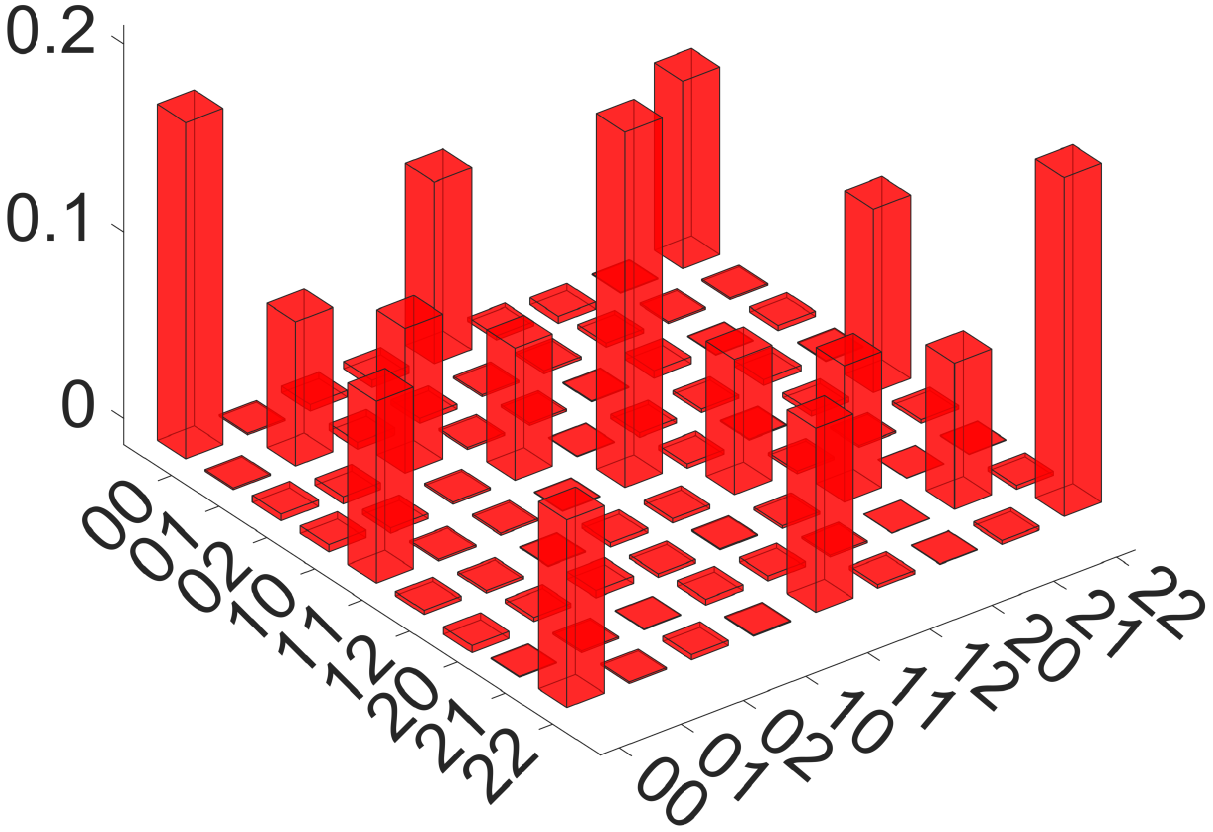}}
		\subfigure(b){\includegraphics [width=8.35cm,height=5.7cm]{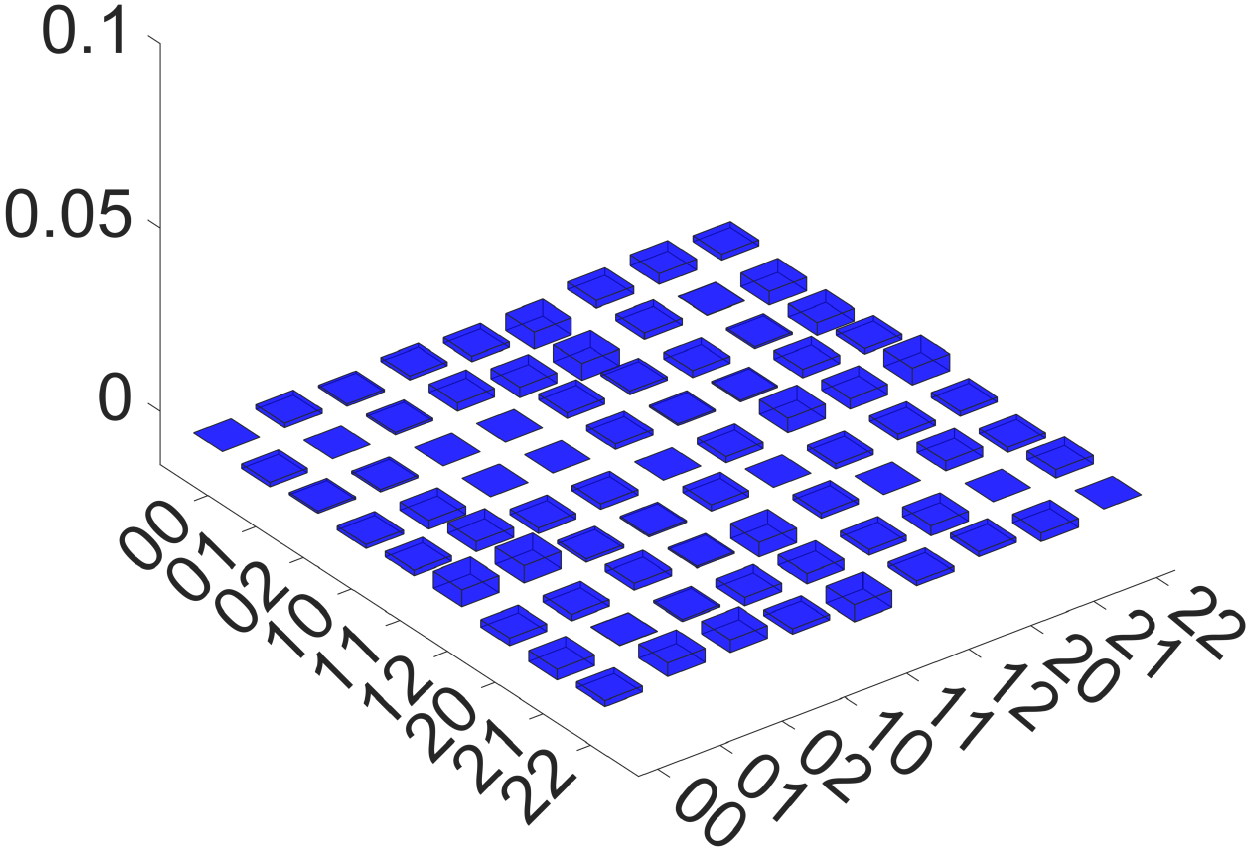}}
	\end{center}
	\caption{Target $v=0.323$, estimate $v=0.303$. (a) The real elements. (b) The imaginary elements.}
	\label{Fig:tomography2}
\end{figure}

\begin{figure}[htbp]
	\begin{center}
		\subfigure(a){\includegraphics [width=8.77cm,height=5.7cm]{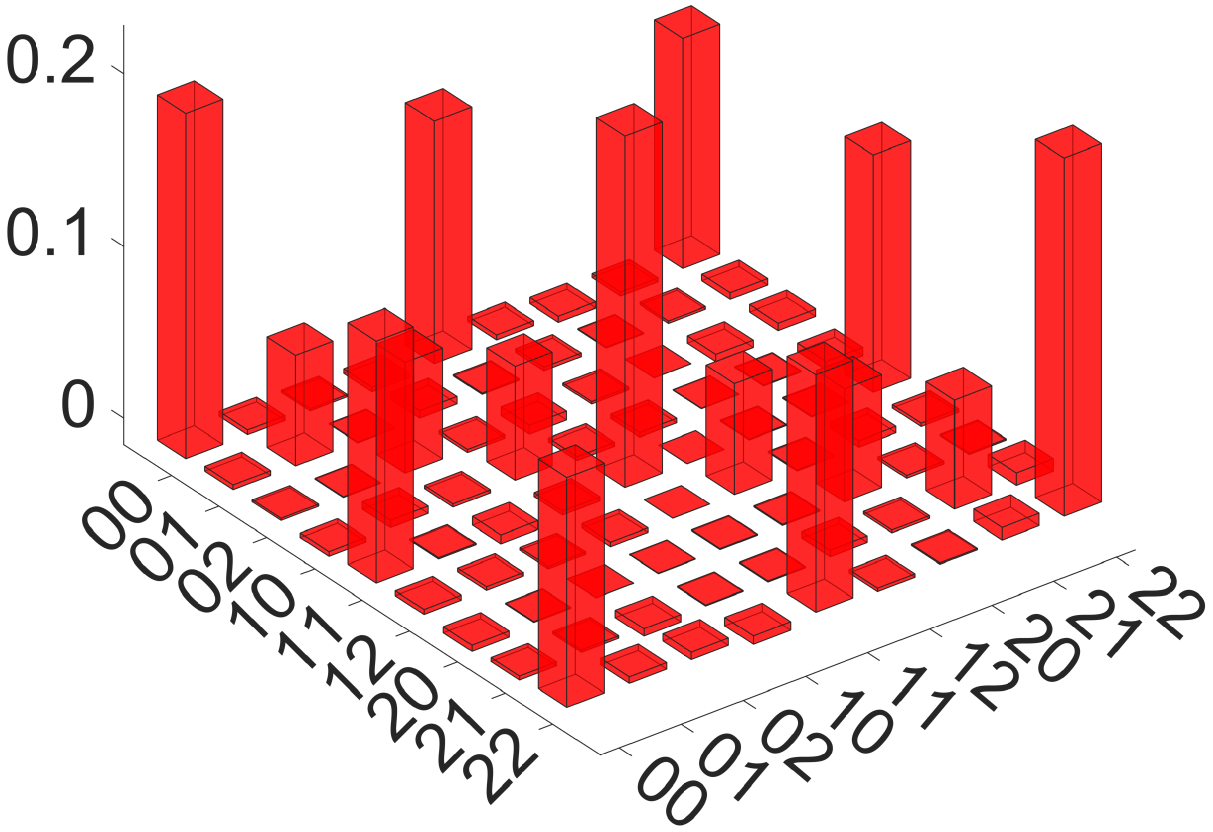}}
		\subfigure(b){\includegraphics [width=8.35cm,height=5.7cm]{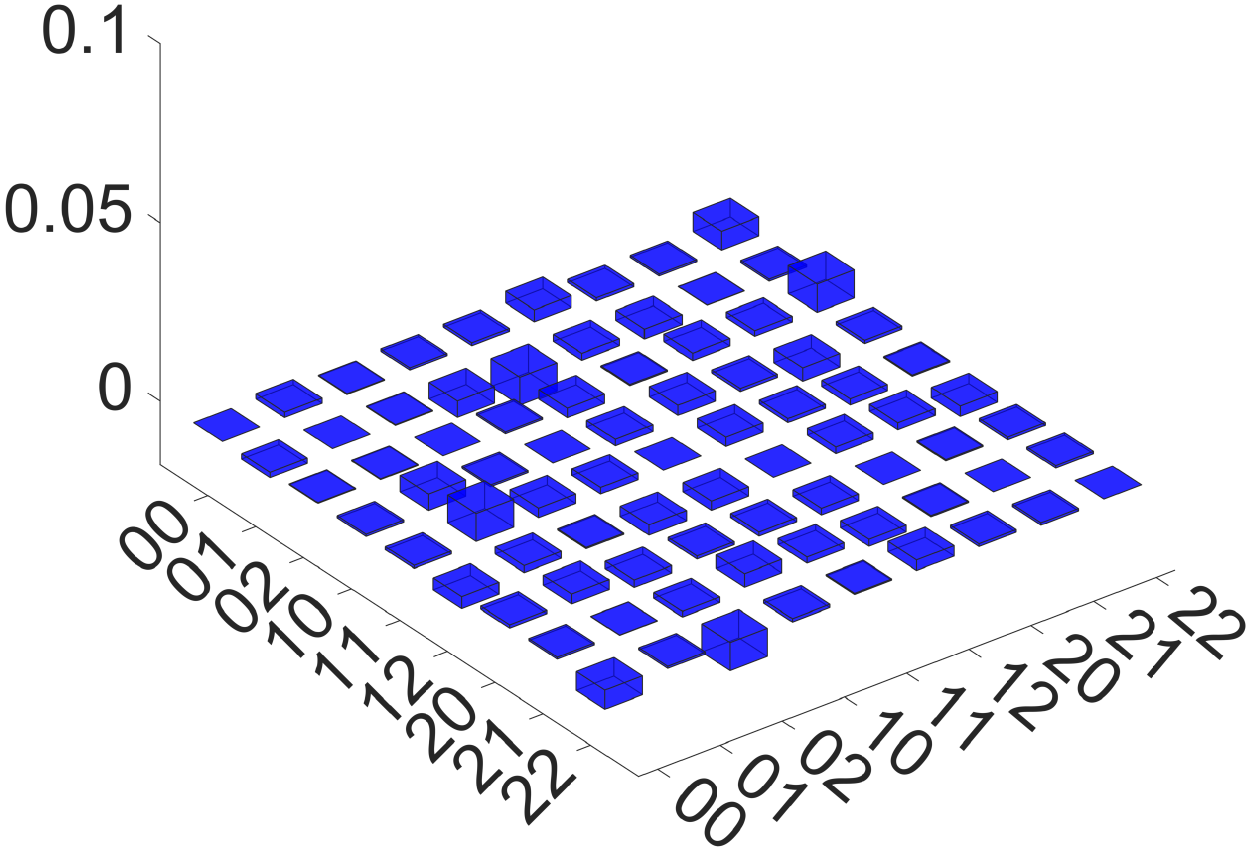}}
	\end{center}
	\caption{Target $v=0.427$, estimate $v=0.413$. (a) The real elements. (b) The imaginary elements.}
	\label{Fig:tomography3}
\end{figure}

\begin{figure}[htbp]
	\begin{center}
		\subfigure(a){\includegraphics [width=8.77cm,height=5.7cm]{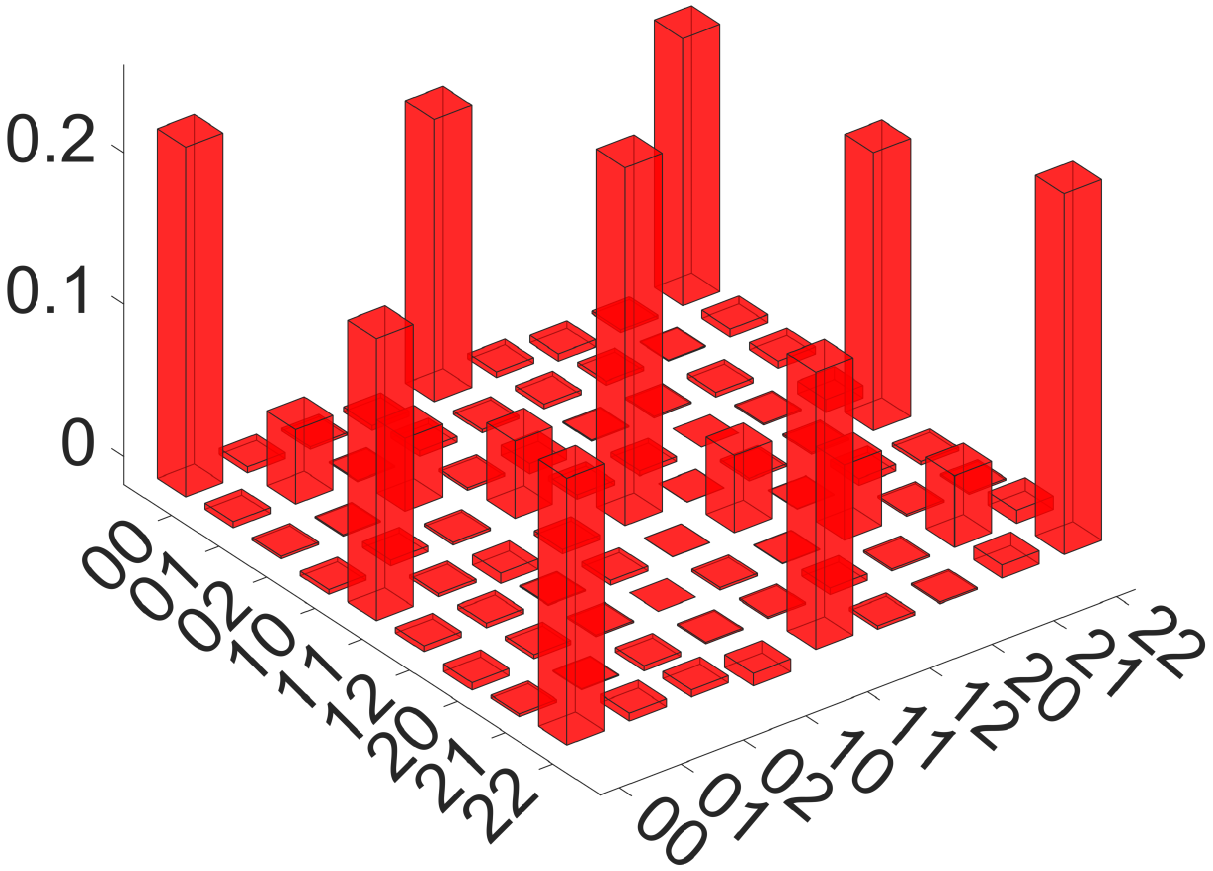}}
		\subfigure(b){\includegraphics [width=8.35cm,height=5.7cm]{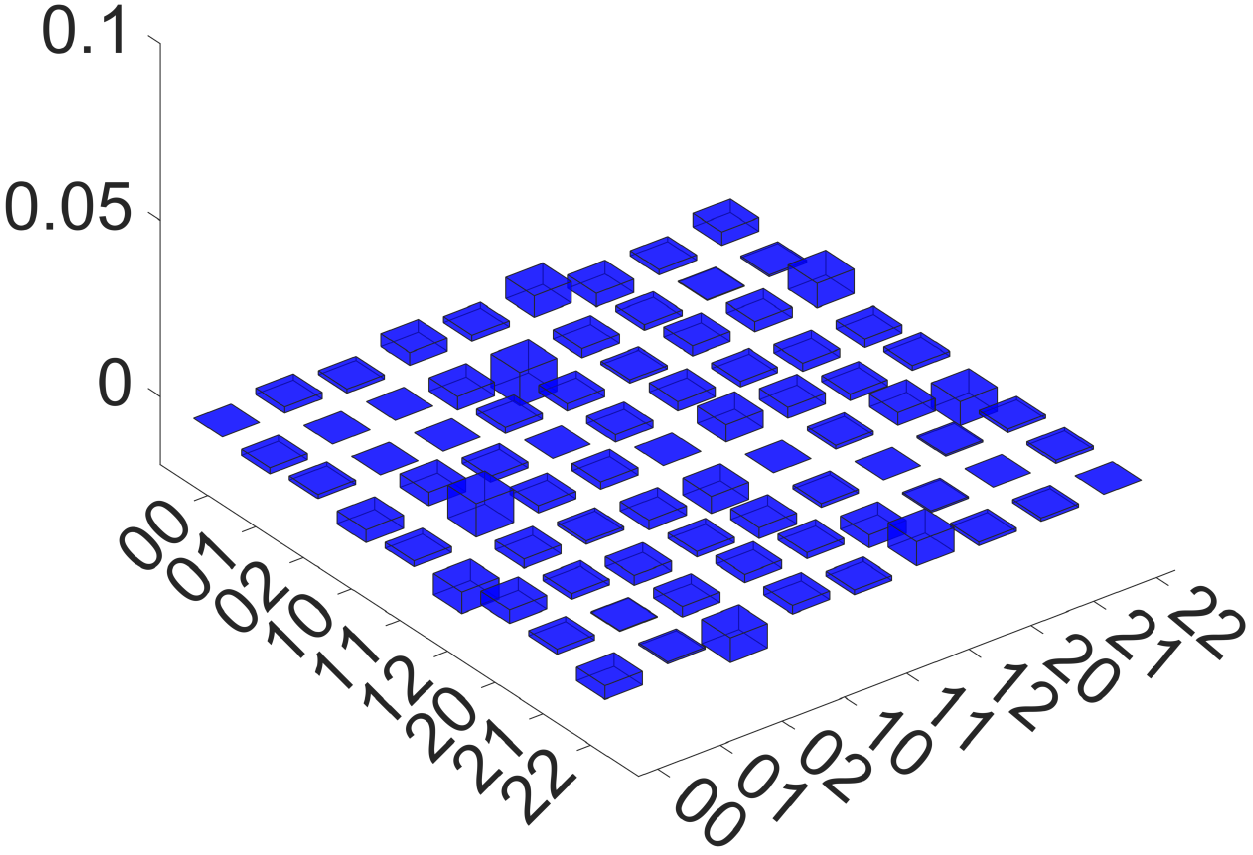}}
	\end{center}
	\caption{Target $v=0.567$, estimate $v=0.547$. (a) The real elements. (b) The imaginary elements.}
	\label{Fig:tomography4}
\end{figure}

\begin{figure}[htbp]
	\begin{center}
		\subfigure(a){\includegraphics [width=8.77cm,height=5.7cm]{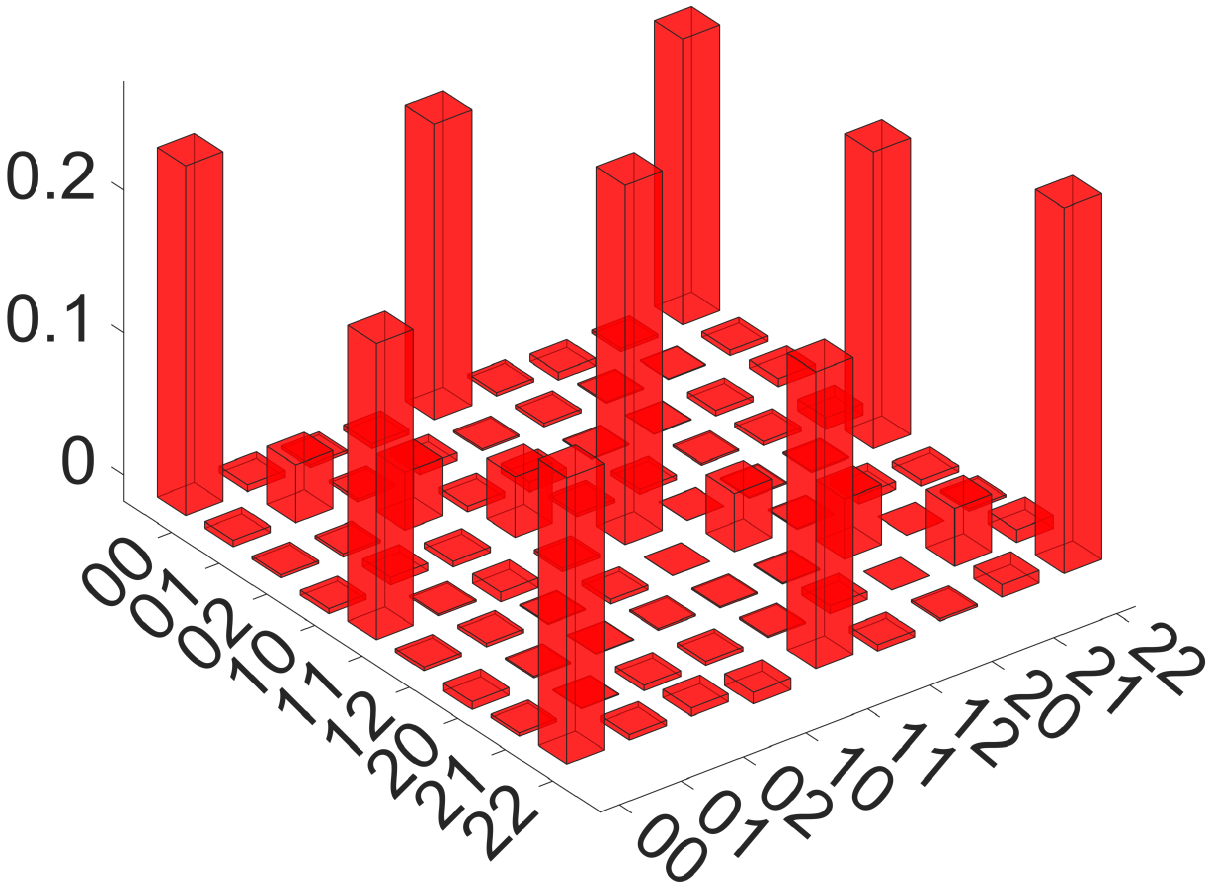}}
		\subfigure(b){\includegraphics [width=8.35cm,height=5.7cm]{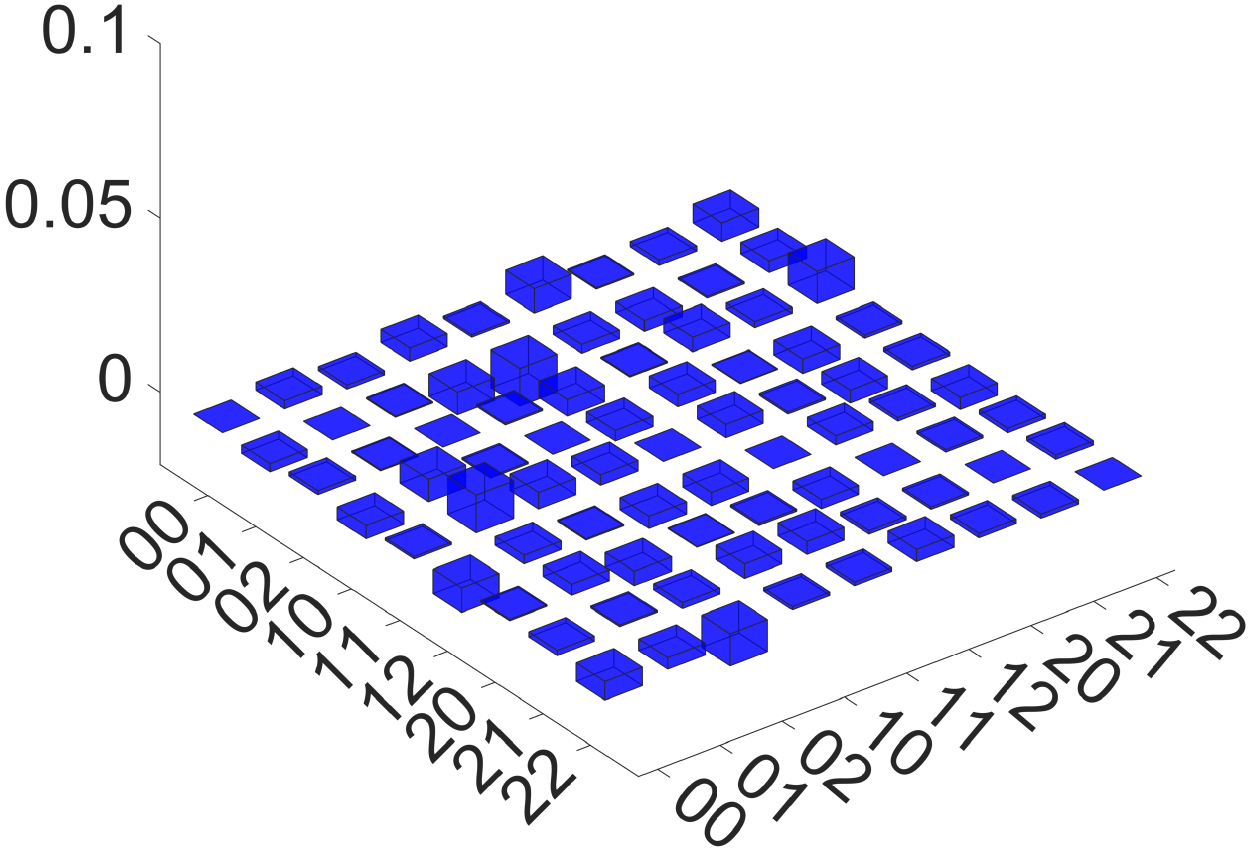}}
	\end{center}
	\caption{Target $v=0.639$, estimate $v=0.621$. (a) The real elements. (b) The imaginary elements.}
	\label{Fig:tomography5}
\end{figure}

\begin{figure}[htbp]
	\begin{center}
		\subfigure(a){\includegraphics [width=8.77cm,height=5.7cm]{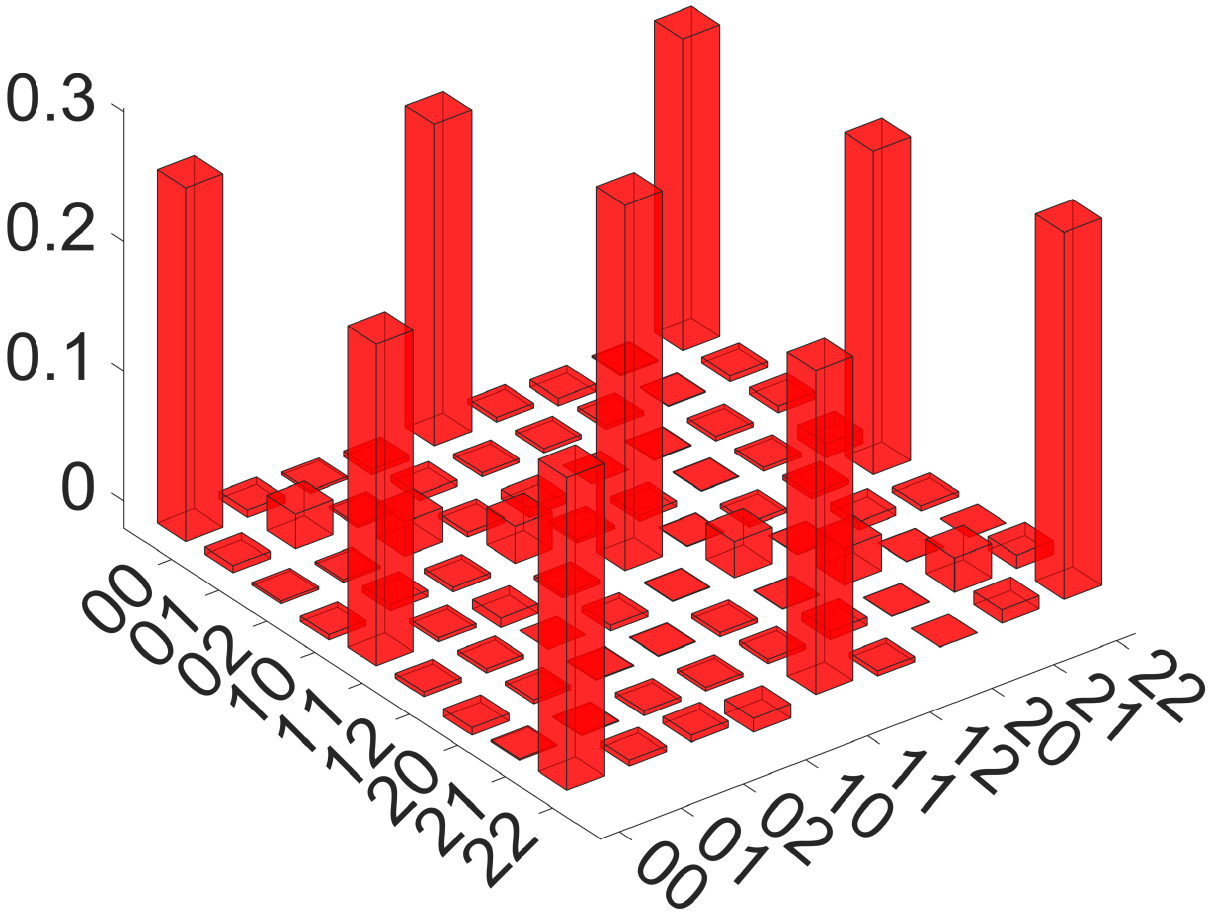}}
		\subfigure(b){\includegraphics [width=8.35cm,height=5.7cm]{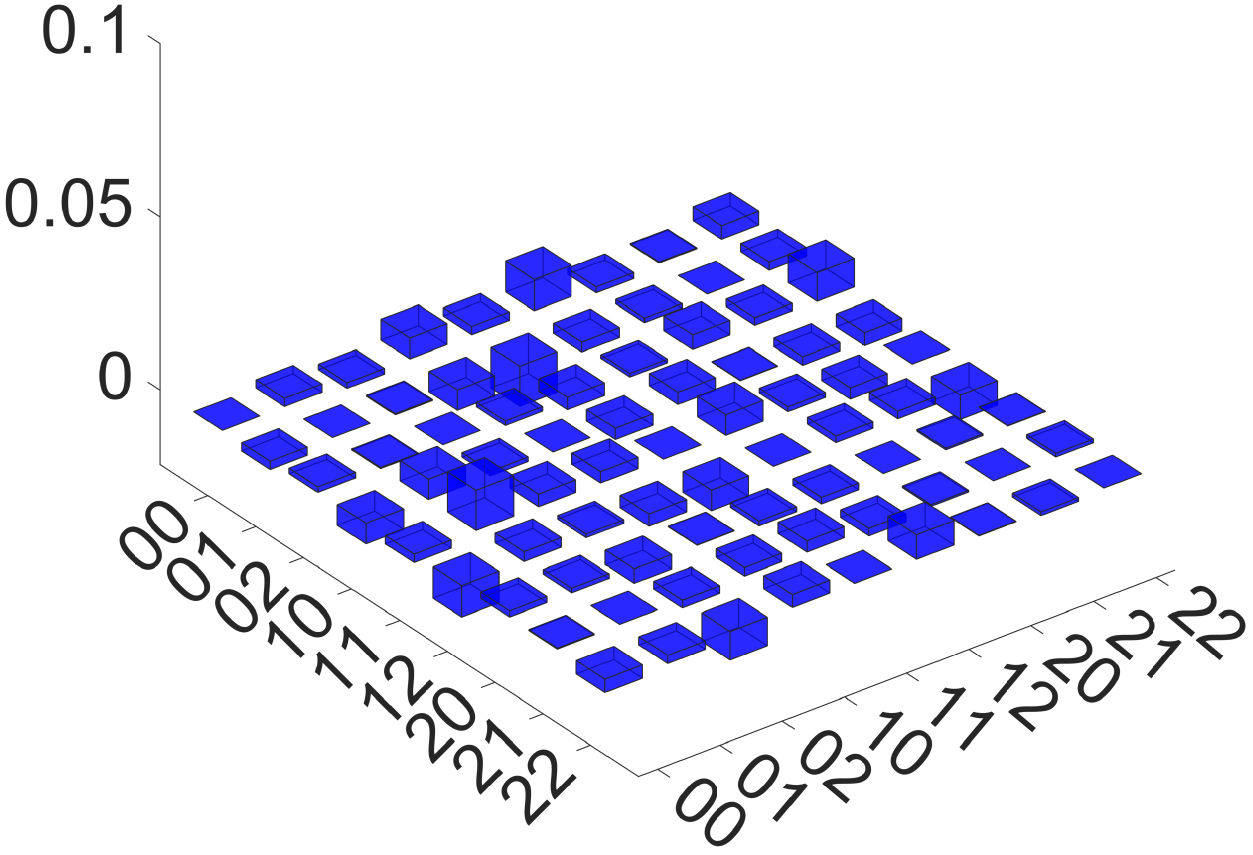}}
	\end{center}
	\caption{Target $v=0.765$, estimate $v=0.742$. (a) The real elements. (b) The imaginary elements.}
	\label{Fig:tomography6}
\end{figure}

\begin{figure}[htbp]
	\begin{center}
		\subfigure(a){\includegraphics [width=8.77cm,height=5.7cm]{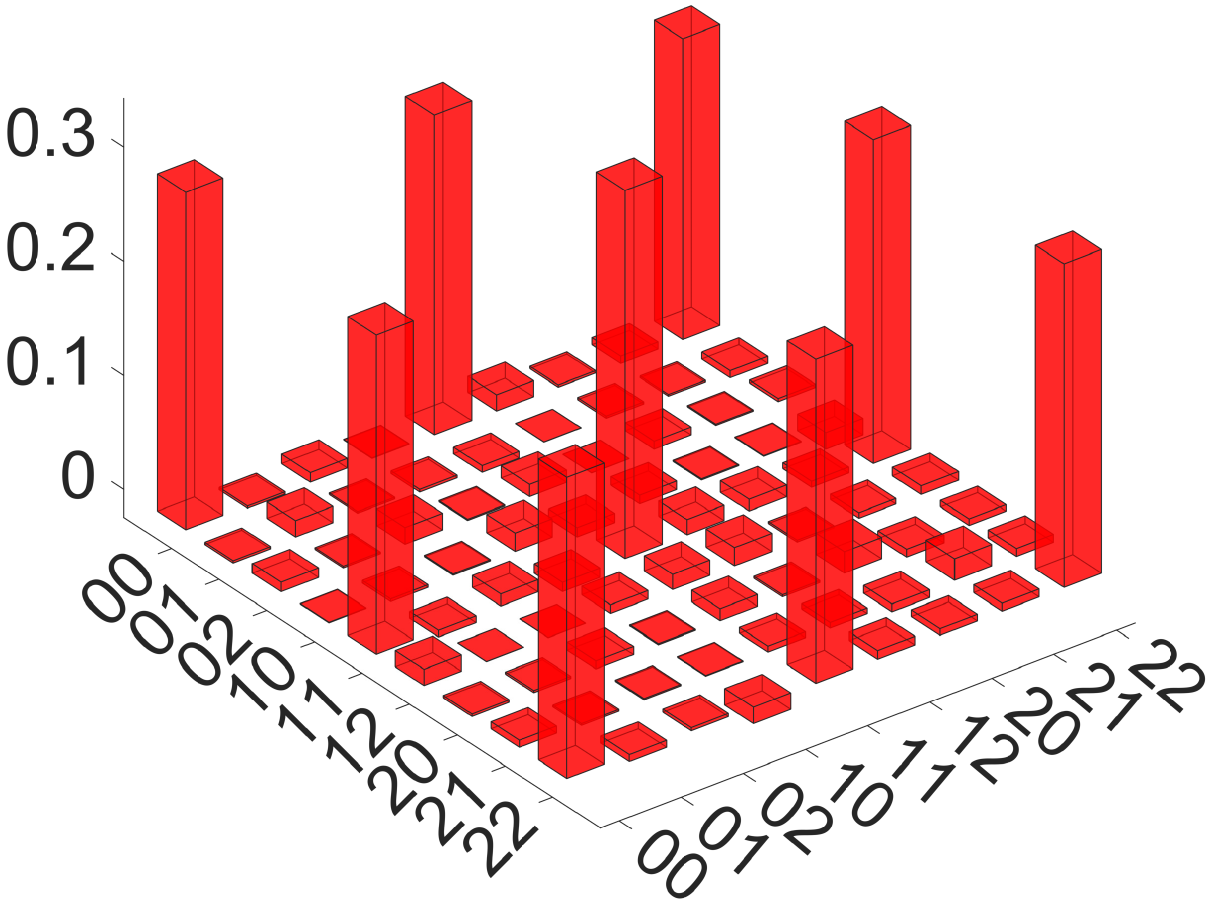}}
		\subfigure(b){\includegraphics [width=8.35cm,height=5.7cm]{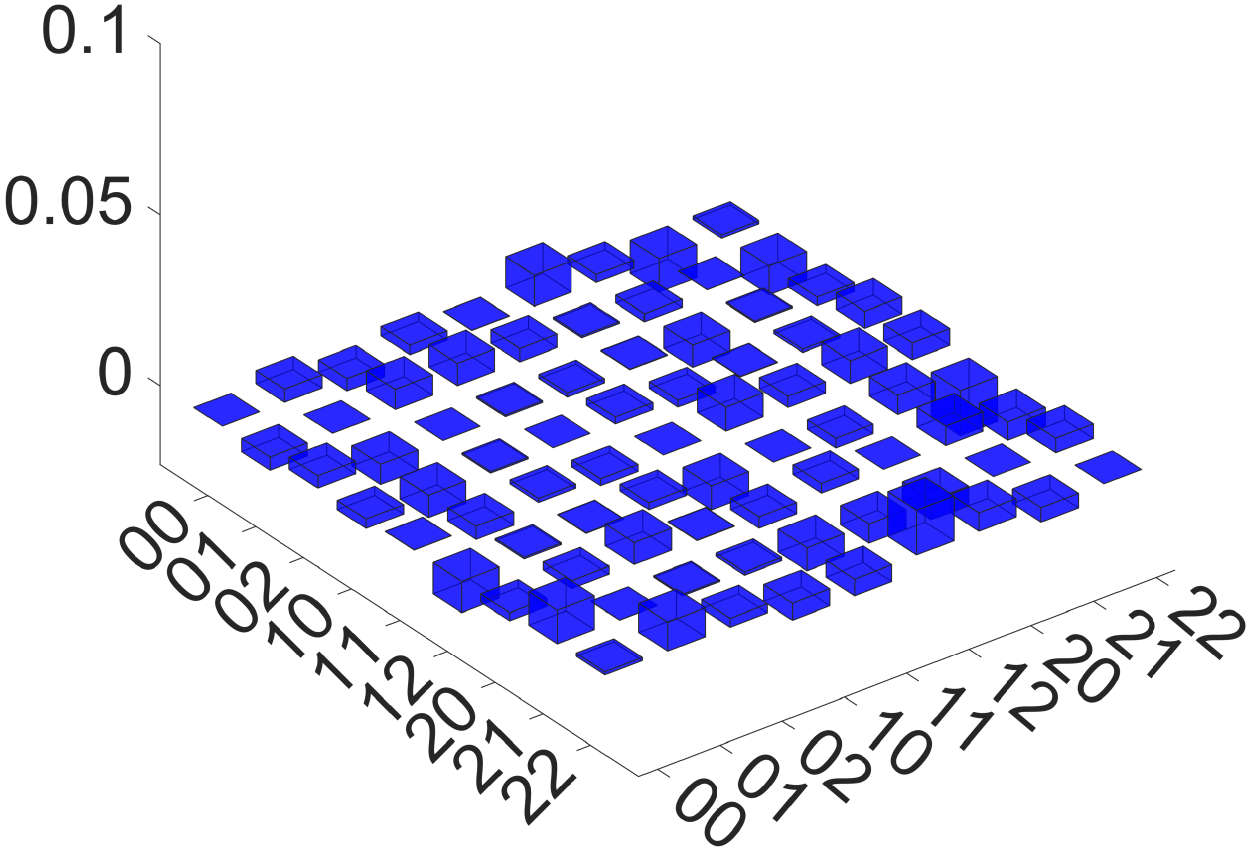}}
	\end{center}
	\caption{Target $v=0.861$, estimate $v=0.833$. (a) The real elements. (b) The imaginary elements.}
	\label{Fig:tomography7}
\end{figure}

\begin{figure}[htbp]
	\begin{center}
		\subfigure(a){\includegraphics [width=8.77cm,height=5.7cm]{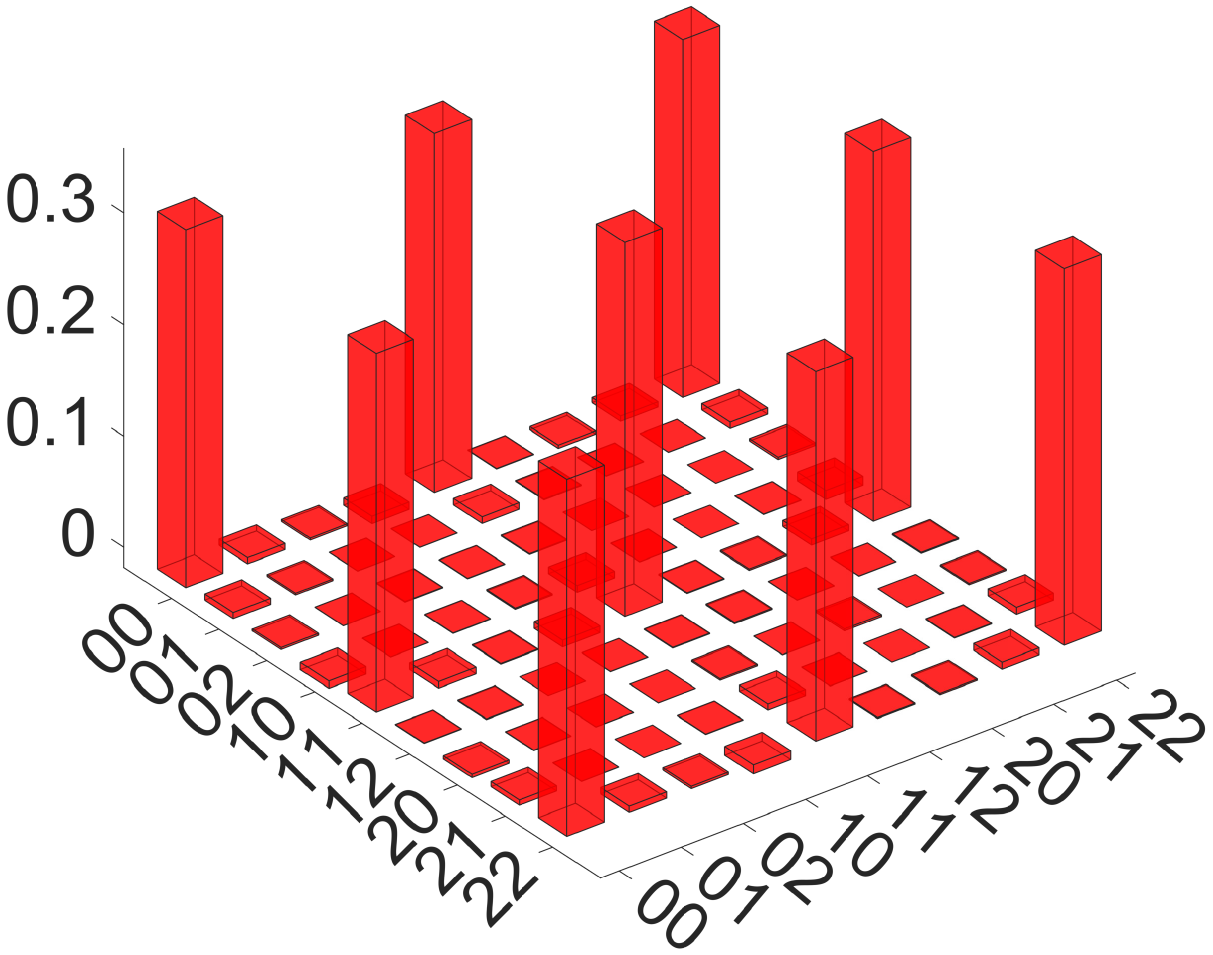}}
		\subfigure(b){\includegraphics [width=8.35cm,height=5.7cm]{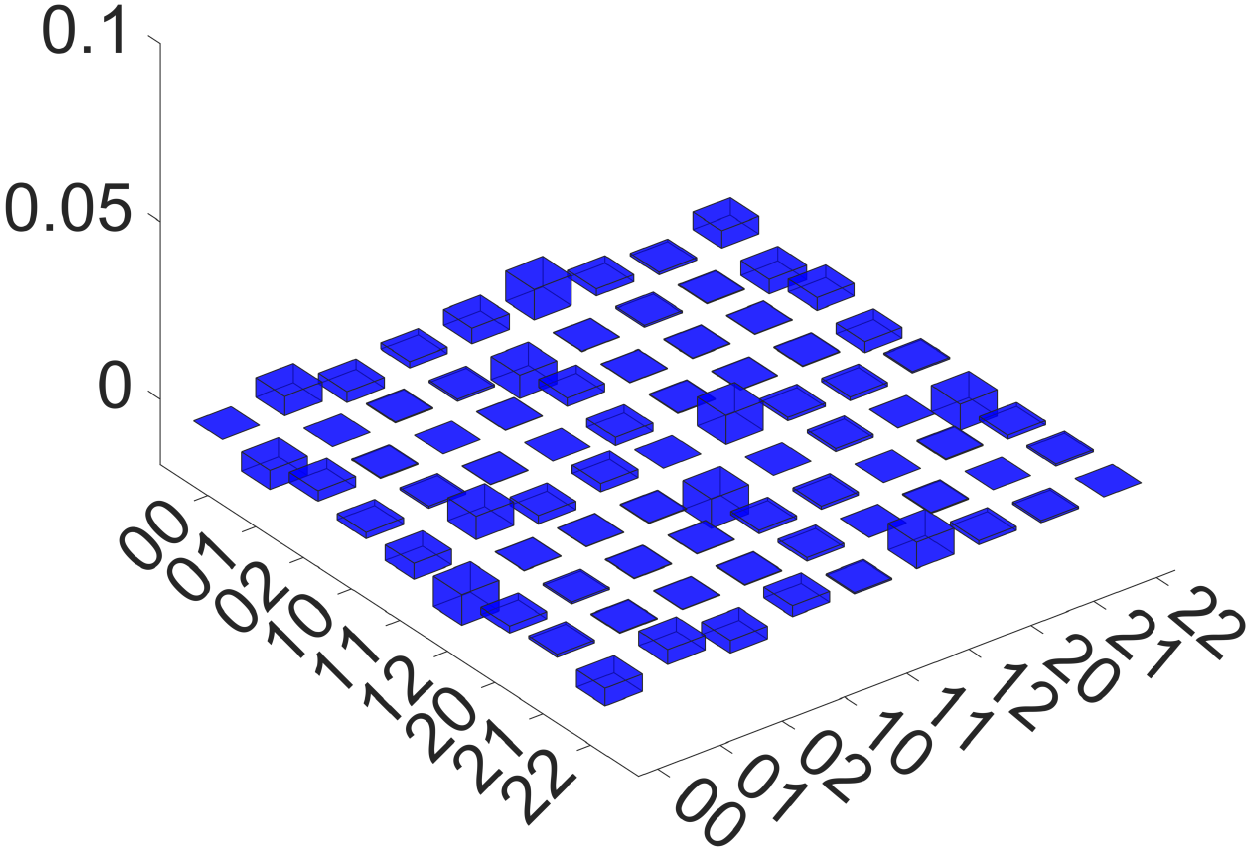}}
	\end{center}
	\caption{Target $v=1$, estimate $v=0.977$. (a) The real elements. (b) The imaginary elements.}
	\label{Fig:tomography8}
\end{figure}

\newpage
\section{Experimental witness values}\label{Appdata}
In Table~\ref{BELLdata} we present the measured values of $\mathcal{B}$ for the eight different visibilities, corresponding to the results Figure in the main text.

\begin{table}[htbp]
	\centering
	\setlength{\tabcolsep}{0.5mm}{
		\begin{tabular}{|c|c|c|c|c|c|c|c|c|}
			\hline
			Target visibility & 0.205 & 0.323 & 0.427 & 0.567 & 0.639 & 0.765 & 0.861 & 1 \\
			\hline
			Estimated visibility & 0.197 & 0.303 & 0.413 & 0.547 & 0.621 & 0.742 & 0.833 & 0.977 \\
			\hline
			$\mathcal{B}$ & -0.418 & 0.438 & 1.304 & 2.375 & 2.985 & 3.942 & 4.667 & 5.796 \\
			\hline
			$\sigma$    & 0.012 & 0.012 & 0.012 & 0.013 & 0.013 &  0.014 & 0.015 & 0.015 \\
			\hline
	\end{tabular}}
	\caption{The experimental values and standard deviations for $\mathcal{B}$ of eight different visibilities}
	\label{BELLdata}
\end{table}%

\end{document}